\documentclass[12pt]{article}%
\usepackage{amsmath,amssymb,amsthm,amsfonts}
\usepackage{wasysym}
\usepackage{graphicx}
\usepackage[dvipsnames]{xcolor}
\usepackage{stackengine}

\usepackage[colorlinks]{hyperref}
\usepackage{tikz}
\usepackage[export]{adjustbox}

\usepackage{geometry}
\usepackage{appendix}

\renewcommand{\epsilon}{\varepsilon}

\newcommand{\del}{\nabla}

\definecolor{red}{rgb}{0.8500, 0.3250, 0.0980}
\definecolor{green}{rgb}{0.4660, 0.6740, 0.1880}
\definecolor{yellow}{rgb}{0.9290, 0.6940, 0.1250}
\definecolor{blue}{rgb}{0, 0.4470, 0.7410}

\begin{document}

\title{Spatio-temporal models of infectious disease with high rates of asymptomatic transmission}

\author{Aminur Rahman \thanks{Corresponding,  \url{arahman2@uw.edu},  \url{http://faculty.uw.edu/arahman2}}
\thanks{Department of Applied Mathematics, University of Washington}, Angela Peace\thanks{Department of Mathematics and Statistics, Texas Tech University, Lubbock, TX}, Ramesh Kesawan\thanks{Department of Statistics, University of Nebraska - Lincoln}, Souparno Ghosh\footnotemark[4]}

\date{}

\maketitle

\begin{abstract}
The surprisingly mercurial Covid-19 pandemic has highlighted the need to not only accelerate research on infectious disease, but to also study them using novel techniques and perspectives.  A major contributor to the difficulty of containing the current pandemic is due to the highly asymptomatic nature of the disease.  In this investigation, we develop a modeling framework to study the spatio-temporal evolution of diseases with high rates of asymptomatic transmission, and we apply this framework to a hypothetical country with mathematically tractable geography; namely, square counties uniformly organized into a rectangle.  We first derive a model for the temporal dynamics of susceptible, infected, and recovered populations, which is applied at the county level.  Next we use likelihood-based parameter estimation to derive temporally varying disease transmission parameters on the state-wide level.  While these two methods give us some spatial structure and show the effects of behavioral and policy changes, they miss the evolution of hot zones that have caused significant difficulties in resource allocation during the current pandemic.  It is evident that the distribution of cases will not be stagnantly based on the population density, as with many other diseases, but will continuously evolve.  We model this as a diffusive process where the diffusivity is spatially varying based on the population distribution, and temporally varying based on the current number of simulated asymptomatic cases.  With this final addition coupled to the SIR model with temporally varying transmission parameters, we capture the evolution of ``hot zones'' in our hypothetical setup.
\end{abstract}

\bigskip
\bigskip

\section{Introduction}
\label{Sec: Intro}

Similar phenomena often appear across many different branches of Science.  Infectious disease spreads within a population through probabilistic interactions, and it has been shown that encounters between organisms may be modeled as finite-speed Brownian-like motion \cite{Gurarie2013EncounterRates}.  Spatially discrete interactions between patches of populations have also been modeled \cite{lloyd2004spatiotemporal} using a Susceptible, Infected, Recovered (SIR) model within a patch.  However, in addition to local Brownian-like motion and interactions between population groups, humans travel long distances, which has its own transport dynamics as shown by Brockmann \textit{et al.} \cite{Brockmann2006}.  Similar to Statistical Mechanics \cite{StatisticalMechanics}, these interactions appear random at the scale of a few agents, but average out at the scale of an entire population.

Throughout the Covid-19 pandemic there was an emphasis on modeling the averaged spread of infections at either the federal or state levels.  However, it is evident that in the beginning of the pandemic resources needed to be distributed to the areas that were going to be hit the hardest as it went from one hot zone to the next.  This would have required fast fine-grained spatio-temporal predictions.  However, for current spatio-temporal models \cite{lloyd2004spatiotemporal}, the computational expense would impede fast predictions.  Using statistical techniques may mitigate much of the computational expense while still providing fine-grained predictions.  Using Brownian motion, for example, we have temporal dynamics that provide an averaged picture of the transport, and spatial distributions, which can be represented using a probability density function.  Taking the product provides spatio-temporal dynamics within the system without having to keep track of each molecule.  Similarly, for infectious disease we have temporal dynamics that can be informed by an SIR-type model, and spatial distributions, which naturally has a probability density function associated with it.

A major issue with SARS CoV2 is the high rates of asymptomatic transmission.  This is due to its unusually long latency period and the likelihood of an asymptomatic case.  If a disease is highly symptomatic, we expect the carrier would know that they have the disease thereby voluntarily limiting contact with others, and they may also be unwell enough to involuntarily reduce contact with others.  On the other hand, with highly asymptomatic diseases, even if someone intends to limit contact if they were to ever get infected, they may not even know that they were infected.  This creates an environment conducive to spatio-temporal transport rather than isolated community spread.

Any complex phenomenon is difficult to model, and the complications present in the spread of SARS CoV2 make the Covid-19 pandemic particularly challenging to model \cite{Adam2020, BertozziChallenges20}.  This manuscript presents new modeling techniques inspired by the study of Statistical Mechanics, and provides preliminary results as motivation for the use of these techniques on infectious disease, especially those with high rates of asymptomatic transmission.  The remainder of the paper is organized as follows:  starting in Sec. \ref{Sec: SIR} we derive the SIR-like model for the county-level temporal dynamics.  Next in Sec. \ref{Sec: Likelihood} we conduct a likelihood-based parameter estimation of the transmission parameters from the model in Sec. \ref{Sec: SIR}.  This reveals how the evolution landscape can change with behavioral or policy decisions such as masking and lockdowns.  Finally, section \ref{Sec: PDE} captures the evolution of ``hot zones''.  We model the redistribution of the simulated asymptomatic cases from the SIR model in Sec. \ref{Sec: SIR} as a diffusive process that is dependent on the population distribution and current number of asymptomatic cases in each county.  In Sec. \ref{Sec: Conclusion} we conclude with a discussion on the modeling framework and promising avenues for future work.

\section{County-wise epidemic temporal dynamics}
\label{Sec: SIR}

Let us first derive an SIR-type model of viral transmission of a highly asymptomatic disease. Consider dividing the population into subpopulations of susceptible individuals $S$, latently infected individuals who aren't yet infectious $L$, infectious individuals who are asympotimatic $I_A$ and symptomatic $I_S$, and recovered individuals with temporary immunity $R$. The base model has the following structure:
\begin{subequations}
 \begin{align}
\frac{dS}{dt} &=\underbrace{-\beta_A SI_A}_{\text{\parbox{2.5cm}{\centering asymptomatic \\[-4pt]  transmission}}} \underbrace{-\beta_S SI_S}_{\text{\parbox{2.5cm}{\centering symptomatic  \\[-4pt] transmission}}}  +\underbrace{\eta R}_{\text{\parbox{1.25cm}{\centering immunity \\[-4pt]loss}}} \label{Model0:eq1}\\
\frac{dL}{dt} &=\underbrace{\beta_A SI_A}_{\text{\parbox{2.25cm}{\centering asymptomatic \\[-4pt]  transmission}}} +\underbrace{\beta_S S^I_S}_{\text{\parbox{2.25cm}{\centering symptomatic  \\[-4pt] transmission}}} -\underbrace{\epsilon L}_{\text{\parbox{1.5cm}{\centering latent get \\[-4pt]  infectious}}} \label{Model0:eq2}\\
\frac{dI_A}{dt} &=\underbrace{\rho\epsilon L}_{\text{\parbox{2.25cm}{\centering latent get \\[-4pt]  asym. infectious}}} -\underbrace{\gamma_AI_A}_{\text{\parbox{1.5cm}{\centering recovery}}}-\underbrace{\mu_AI_A} _{\text{\parbox{2cm}{\centering  disease induced \\[-4pt] mortality}}}  \\
\frac{dI_S}{dt} &=\underbrace{(1-\rho)\epsilon L}_{\text{\parbox{2.5cm}{\centering latent get \\[-4pt]  sym. infectious}}} -\underbrace{\gamma_SI_S}_{\text{\parbox{1.5cm}{\centering recovery}}}-\underbrace{\mu_SI_S} _{\text{\parbox{2cm}{\centering  disease induced \\[-4pt] mortality}}}  \\
\frac{dR}{dt} &=\underbrace{\gamma_AI_A+\gamma_SI_S}_{\text{\parbox{3cm}{\centering recovery}}} -\underbrace{\eta R}_{\text{\parbox{2.5cm}{\centering immunity loss}}}
 \end{align}
\label{Eq: Model0}
 \end{subequations}
where $\beta_A$ and $\beta_S$ are the asymptomatic and symptomatic transmission rates respectively, $\epsilon^{-1}$ is the incubation period, $\rho$ is the percentage of cases that are asymptomatic, $\gamma_A$ and $\gamma_S$ are the asymptomatic and symptomatic recovery rates, $\mu_A$ and $\mu_S$ are the disease induced mortality rates for asymptomatic and symptomatic cases, and $\eta$ is the waning rate of the temporary immunity gained from recovery.

While the disease discussed in this investigation is hypothetical, we will, whenever possible, use features of the Covid-19 pandemic to inform parameters and modeling strategies.  The results however, are intended to be proof-of-concept demonstrations of this hypothetical disease and not predictors of Covid-19 dynamics.

We often think of infectious disease parameters as stagnant -- a property of the disease itself.  However, the environment in which it spreads can have a significant effect on these parameters as we have noticed with the current pandemic. Consequently, specifying time-varying parameters make the model more flexible and perhaps more realistic. However, we do not have sufficiently resolved observational information that would enable us to infer the evolution of the said parameters. Hence, we need to fix some of these parameters and infer about the dynamics of the rest. In the current study, we assume disease recovery rates ($\gamma_s$ and $\gamma_a$) and percentage of asymptomatic cases ($\rho$) are constants.  As such, we specify $\gamma_s = 1/25.2$ and $\gamma_a = 1/22.6$ \cite{CovidViralShedding}, and $\rho = 0.25$ \cite{CovidViralShedding, CovidMetaAnalysis}.  The asymptomatic mortality parameter $\mu_A$ is set to zero, and the symptomatic mortality paramter is set to $\mu_S = 0.005\gamma_S$ \cite{UsherwoodCovid2021}.  Further, we let the rate leaving the latent class be $\epsilon = 1/5.8$ \cite{McAloonCovid2020}.  Finally, the time-varying transmission parameters $\beta_S$ and $\beta_A = \rho\beta_S/(1-\rho)$ are estimated using the functional regression technique described in Section \ref{Sec: Likelihood}.

\begin{figure}[htbp]
    \centering
    \stackinset{l}{}{t}{}{\textbf{(a)}}{\includegraphics[width = 0.47\textwidth]{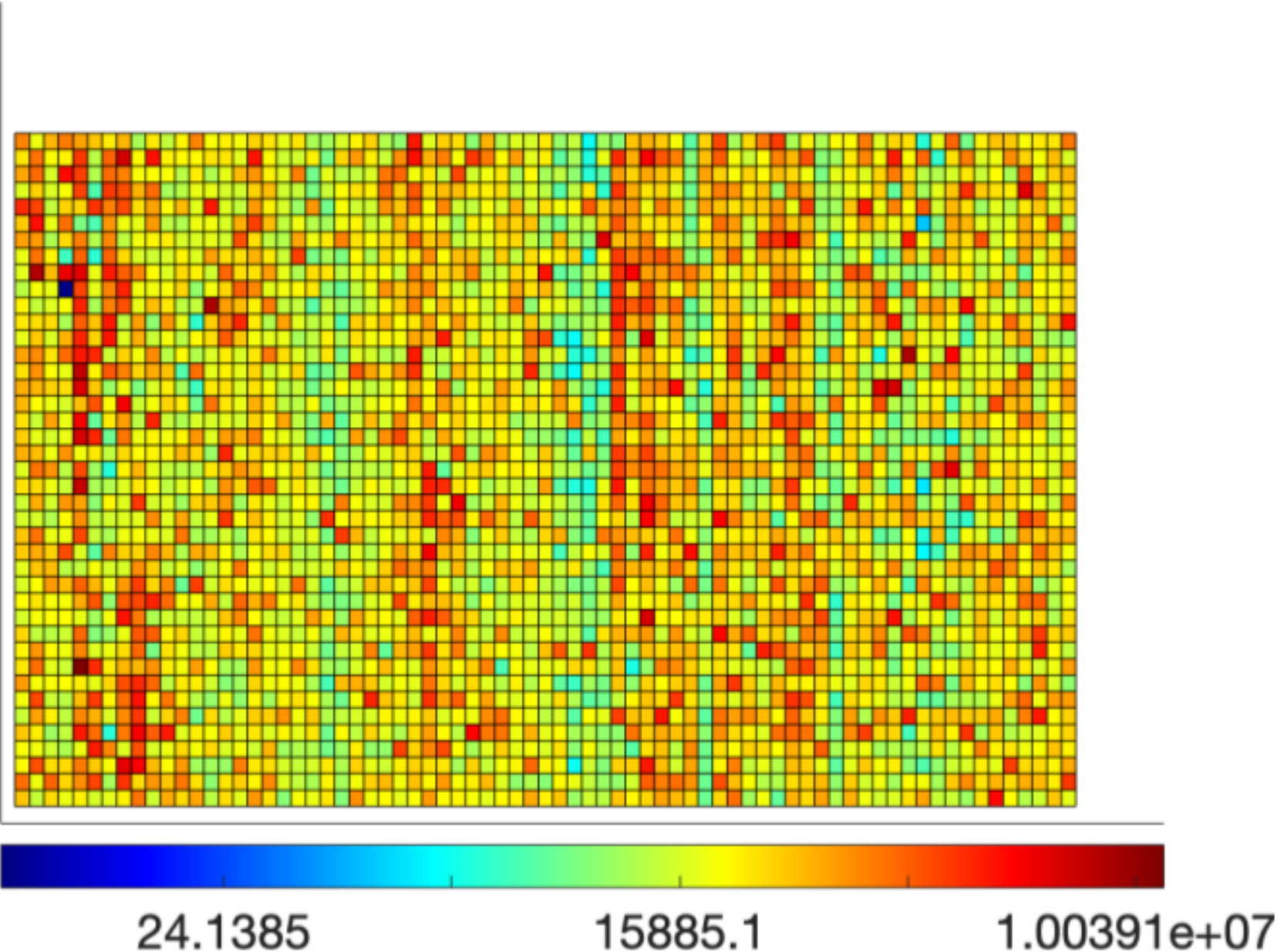}}\qquad
    \stackinset{l}{}{t}{}{\textbf{(b)}}{\includegraphics[width = 0.43\textwidth]{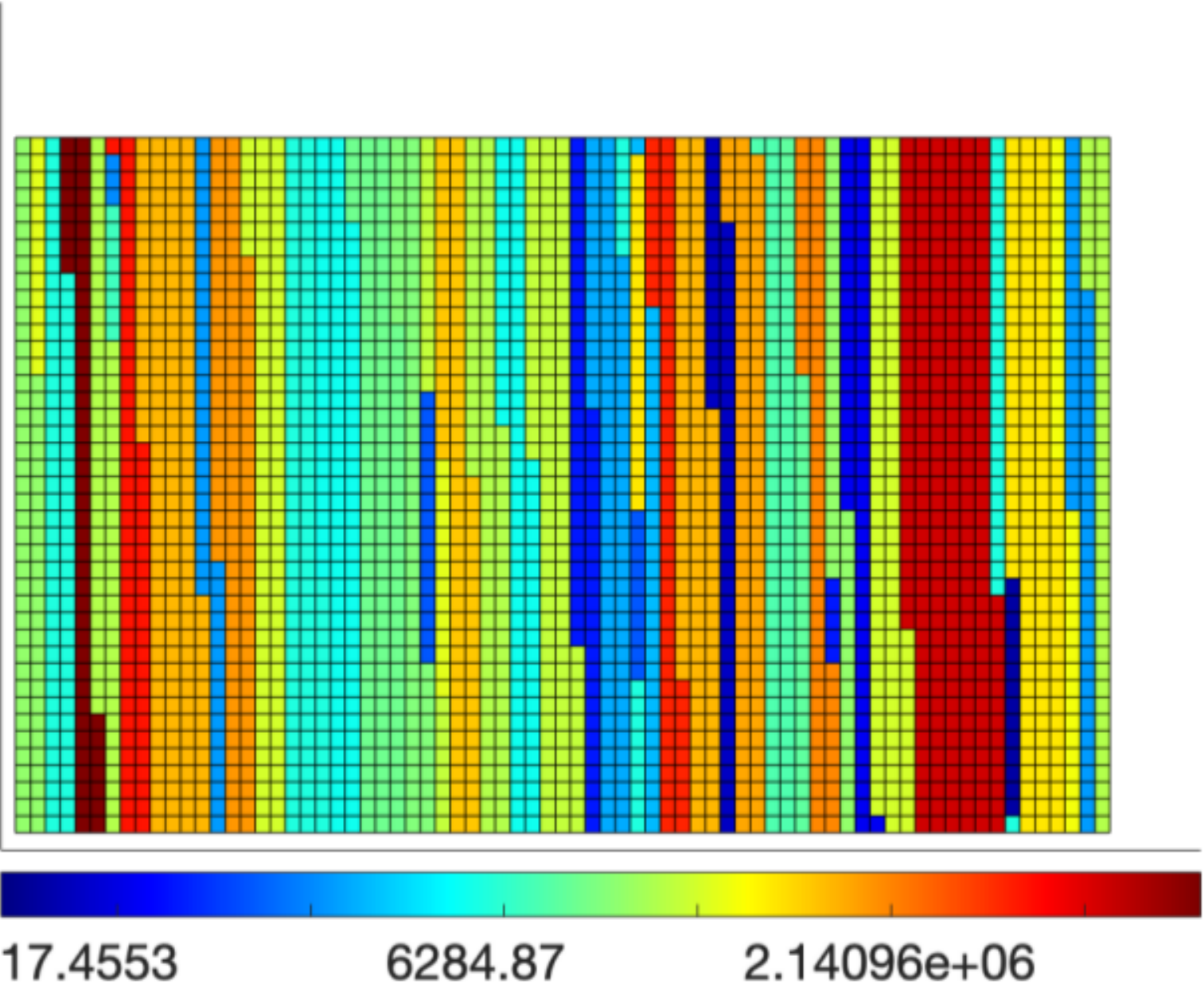}}
    \caption{Population distribution of $\Omega$.  \textbf{(a)} County-wise populations.  \textbf{(b)} State-wide populations.  The colormap, with color bars on the bottom, for the heat maps are on a $\log$-scale for improved visibility.}
    \label{Fig: Population Heat Map}
\end{figure}

Consider a fictional country, let us call it $\Omega$, where all the counties are uniform squares organized into a rectangular grid of $42 \times 74$ forming $48$ contiguous states running sequentially down the columns.  We shall use data from \url{usafacts.org} \cite{USAFacts} for the population and initial conditions for each county.  The county-wise and state-wide population distribution of $\Omega$ is shown in Fig. \ref{Fig: Population Heat Map}\textbf{(a)} and \textbf{(b)}, respectively.  In the heat maps the data for the states are taken in alphabetical order from left to right with each county represented down the column.  So the $\Omega$ state that uses data from California will take up multiple columns, whereas the equivalent of Rhode Island will only take up three rows of a column.  As a more concrete example, we can see California in deep red towards the left of the heat map, Fig. \ref{Fig: Population Heat Map} \textbf{(b)}, with its $58$ counties and Texas towards the right with, 
in the words of Beto O'Rourke, ``all 254 counties''.

\begin{figure}[htbp]
\vspace{-0.5in}
    \centering
    \stackinset{l}{56mm}{t}{1pt}{\textbf{(a)}}{\stackinset{r}{56mm}{t}{1pt}{\textbf{(b)}}{\stackinset{l}{56mm}{b}{51mm}{\textbf{(c)}}{\stackinset{r}{56mm}{b}{51mm}{\textbf{(d)}}{\includegraphics[width = 0.9\textwidth]{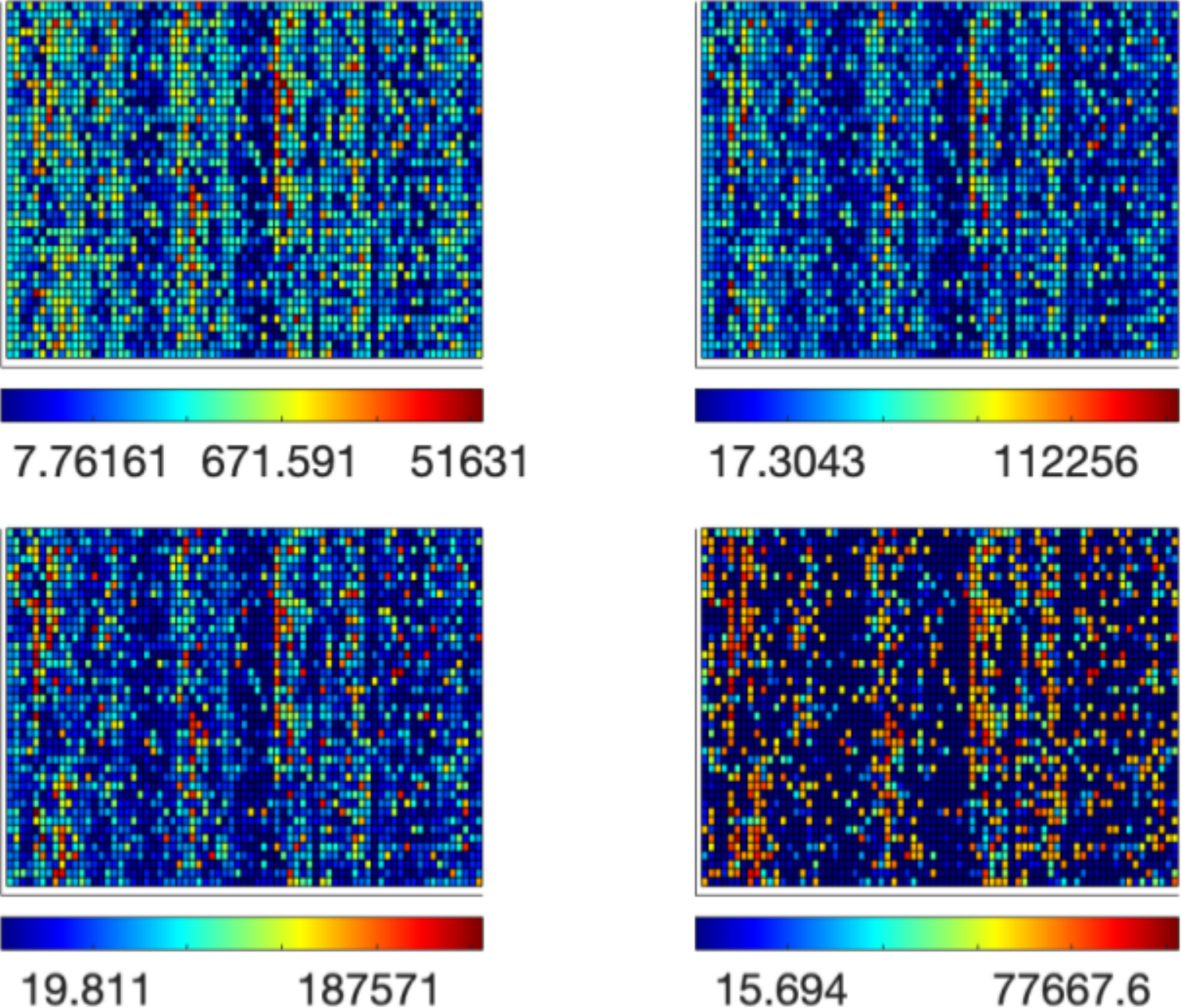}}}}}
    \stackinset{r}{4mm}{t}{7mm}{\textbf{(e)}}{\includegraphics[width = 0.75\textwidth]{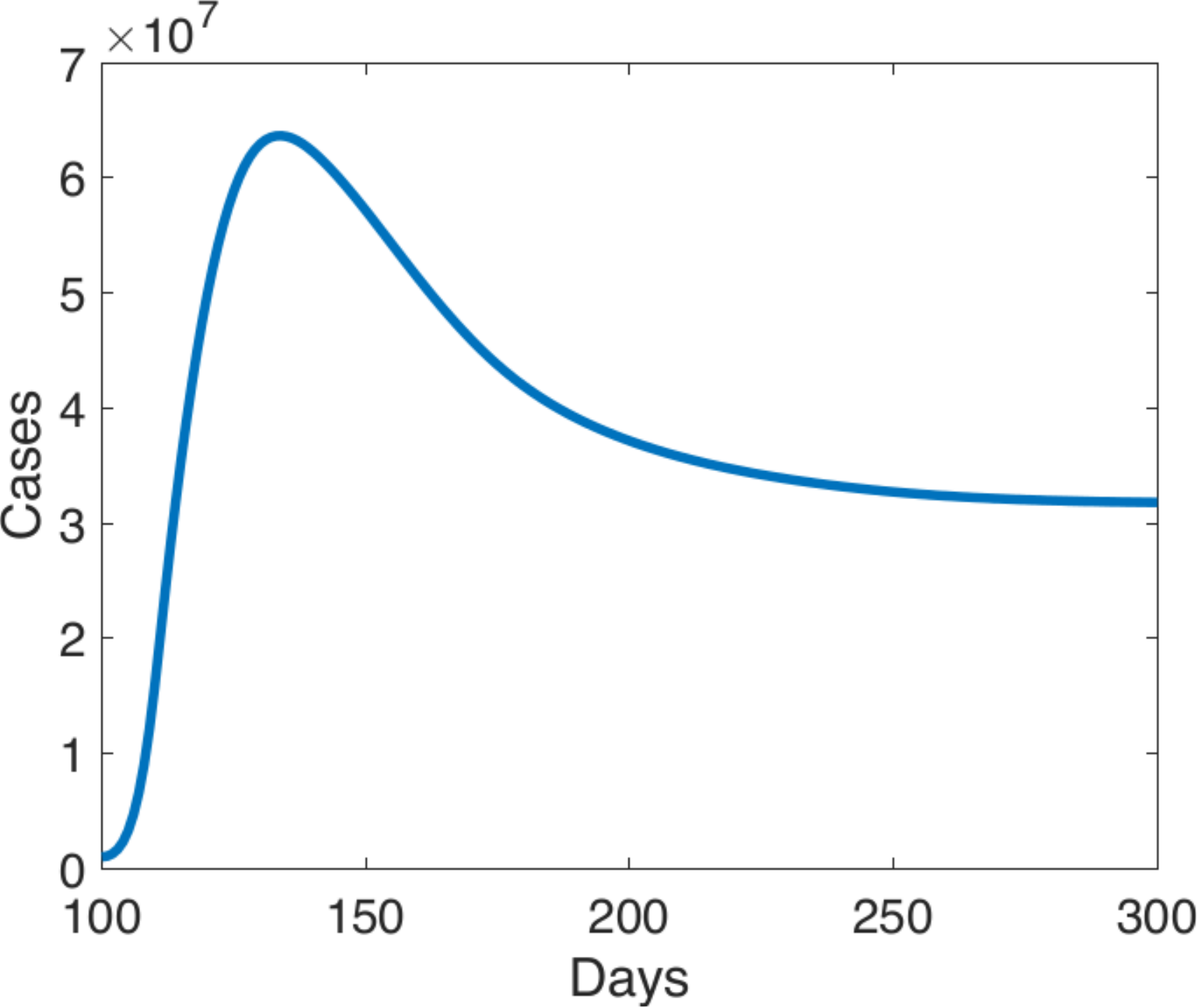}}
    \caption{Simulation of the SIR model \eqref{Eq: Model0} with initial condition \textbf{(a)} taken from the $100$ day mark.  \textbf{(b)} - \textbf{(d)} simulated heat maps at the $110$, $130$, and $300$ day marks.  The colormap, with color bars on the bottom, for the heat maps are on a $\log$-scale.  \textbf{(e)}  Daily case totals for $\Omega$ showing logistic-like growth for the first few weeks, and then a decline to a steady-state.}
    \label{Fig: SIR}
\end{figure}

Using a representative value of $\beta_S = 0.02$ we may simulate the SIR model for self-contained counties; i.e., it is assumed that there there is no interaction between the counties.  Heat maps for the simulation of the number of symptomatically infected individuals, $I_S$, are produced by solving \eqref{Eq: Model0} through the standard MATLAB Runge-Kutta \cite{Runge, Kutta} solver \verb|ode45|.  The heat maps are shown in Fig. \ref{Fig: SIR}\textbf{(a - d)} with the initial condition in \textbf{(a)} calculated at the $100$ day mark from the \url{usafacts.org} data set \cite{USAFacts}.  Fig. \ref{Fig: SIR}\textbf{(b - d)} show the solutions to the SIR model \eqref{Eq: Model0} at day $110$, $130$, and $300$.  The colormap for the heat maps are on a $\log$-scale to highlight the large differences in disease proliferation between counties, which is also what has been observed for the current pandemic.  The daily totals for the entirety of $\Omega$ is shown in Fig. \ref{Fig: SIR}\textbf{(e)} where we observe an initial logistic-like growth eventually decaying to a steady-state.

In addition to the predicted evolution of a disease, epidemiologists often estimate a basic reproduction number, $\mathcal{R}_0$.  With our SIR modeling setup, we can compute the $\mathcal{R}_0$ using the ``next generation matrix approach'' of Diekmann and co-workers \cite{diekmann2010construction}: 
\begin{equation}
    \mathcal{R}_0=\left(\frac{\beta_a\rho}{\gamma_a+\mu_a}+\frac{\beta_s(1-\rho)}{\gamma_s+\mu_s}\right)N
    \label{Eq: R0}
\end{equation}
where $N$ is the normalized population of the region of interest.  Once we estimate the temporal evolution of the transmission parameters $\beta_S$ and $\beta_A$ in Sec. \ref{Sec: Likelihood}, we can estimate the $\mathcal{R}_0$, and investigate how it too evolves over time.

\section{State-wide likelihood-based parameter estimation}
\label{Sec: Likelihood}

With an SIR-type population model and accurately estimated parameters, we would be able to describe the averaged effects of the virus on a community. Standard SIR-type models, like System \eqref{Eq: Model0} can be spatially expanded using spatially discrete patches each with coupled systems of ordinary differential equations (ODEs) \cite{van2008spatial}. However, if we use a full patch modeling approach the dimension of the system can be extremely large as the number of patches is increased, and the populations are further divided into subpopulation compartments within each patch. Additionally, the size of parameter space will also increase as the model dimension increase leading to computational instability, or even intractability.

Thus, to reduce computational complexity some disease specific epidemiological characteristics could be used as auxiliary information. For instance, it is observed that SARS-CoV2 symptomatically infects individuals with some probability that is dependent on age, number of encounters with infected individuals, respiratory health, population density, and proximity to an infection ``hot zone''.  We hypothesize that this individual-level probability manifests itself as a spatial continuum through a spatial stochastic process. Let  $\Psi(x,y)$ be the joint probability density function (pdf) for a finite collection of random variables operating in space.  Applying the pdf locally (near a ``hot zone'') to an averaged symptomatically infected population yields a spatial distribution of the most vulnerable individuals; that is, $u(x,y;t) = \Psi(x,y)I_s(t)$. Observe that, for a given $t$, the formulation of $u(x,y;t)$ is reminiscent of intensity function driving a non-homogeneous spatial point process in the sense that $\int_{(x,y)\in\mathcal{A}} u(x,y,t)dx dy= I_s(t)$\cite{illian2008statistical}. In other words, at time $t$, the spatial process $\Psi(.,.)$ redistributes $I_s(t)$ over a bounded spatial domain $\mathcal{A}$. Of course, these ``hot zones'' expand and/or move over time, and for long time dynamics of a pandemic we may need to introduce non-separability of space and time in $\Psi(.)$, that is, $u(x,y;t) = \Psi_t(x,y)I_s(t)$ where $\Psi_t(.)$ can be constructed using non-separable space-time covariance matrices \cite{bruno2009simple, rodrigues2010class}.  However, even with the short time localized distribution, we may presumably be able to quickly implement policies that mitigates an epidemic or prevents a pandemic and send resources to necessary locations to save as many lives as possible.

Customarily, the transmission rate parameters $\beta_A,\beta_S$ in \eqref{Model0:eq1} are assumed to be constant. However, in reality, they  can change under external forcing (for example: policy intervention). Hence, we propose to statistically estimate a temporal evolution of $\beta_A$ and $\beta_S$, explicitly taking into account a policy intervention event - the implementation of mask mandates in several US states. In estimating $\beta$, we only focus on \eqref{Model0:eq1} because the time series of susceptible population can be observed and therefore an empirical model can be formulated\footnote {The trajectory of $L$ in \eqref{Model0:eq2} is guided by unobserved latent processes and therefore observational models cannot be specified.}. However, fitting a statistical model with the parameterization in \eqref{Model0:eq1} is problematic because $\beta_A$ and $\beta_S$ are not identifiable individually. Hence, we reparameterize the statistical model in the following way: 

Let $S_j(t)$ be the susceptible population of state $j$ at day $t$. Since the observed time series is at daily resolution, we define $Y_j(t):=S_j(t)-S_j(t-1), t=1,2,..,T$ as a discrete approximation of the LHS of \eqref{Model0:eq1}. We now model the time series $Y_j(t)$  using the functional predictors $S^*_j(t-1)$, that quantifies the latent population level evolution of the susceptible individuals and a binary vector $Z_{j,t}$ that takes the value 1  (0) if mask mandate was (was not) in effect at time $t$ in the $j$th state. Although $Z$ is time indexed, it difficult to envision it as a function. Hence given a time point, we simply view $Z_{.,t}$ as fixed predictor available to us at that time point.  

Observe that, for the $j$\textsuperscript{th} state, we view $Y_j(t)$ as a random response function dependent on two predictor functions and therefore can be statistically modeled using the function-on-function regression techniques. Formally, we write the regression model as follows,
\begin{equation}
\label{fcr}
Y_{j}(t)=f_{0}(t)+S^*_j(t-1)\beta(t) +Z_{j,t} f_{1}(t)+b_{j}(t)+\epsilon_{j}(t),
\end{equation}
where $f_{0}(t)$ is the population level functional intercept term, $\beta(t)$ is the slope function modeling the \textit{concurrent} association between $S^*(t-1)$ and $Y(t)$ and therefore can be viewed as the composition of $\beta_A$ and $\beta_S$ in \eqref{Model0:eq1}, $f_1(t)$ quantifies the association between mask mandate and $Y(t)$ at the population, $b_{j}(t)$ is the state-specific random fluctuations from the  $f_{0}(t)$, and $\epsilon_{j}(t)$ are pure noise.

To ensure smoothness $b_{j}(t)$ is modeled as a zero mean Gaussian process with an AR(1) covariance function, $\epsilon_{j}(t)$ are assumed to be independent  $N\left(0, \sigma_{\epsilon}^{2}\right)$ with $b_{j}(t)$ and $\epsilon_{i}\left(t\right)$ are mutually independent across $j$. Furthermore, since model $S^*_{j}\left(t-1\right)$ in the RHS of \eqref{fcr} is a latent process, we propose the following process level specificaiton of $S^*$ process $S_{j,obs}(t)=\mu_{S}\left(t\right)+b_{j,S}\left(t\right)+\epsilon^s_{j}(t)$, with $S^*_j(.)=\mu_{S}\left(.\right)+b_{j,S}\mu_{S}(.)$  and $\epsilon^s_j(.)$ is again a pure noise process.  $\mu_{S}(.)$ the population-level mean of the $S^*$ process  and $b_{j,S}(.)$ is the state-specific deviation from the population mean of the $S^*$ process.The regression coefficient functions are modeled using cubic B-splines with 30 interior knots. Thus, denoting $\mathbf{\chi}(t)$ $=\left\{\chi_{1}(t), \cdots, \chi_{30}(t)\right\}^{\prime}$ as a sequence of B-spline basis functions evaluated at $t$, then $\beta(t)=\mathbf{\chi}(t)^{\prime} \boldsymbol{\theta}$ with  $\boldsymbol{\theta}=\left(\theta_{1}, \cdots, \theta_{30}\right)^{\prime}$ being the vector of coefficients. Similarly, state-specific intercept curves $b_{i}(t)$, are also approximated using cubic B-spline basis functions as $b_{j}(t)=\sum_{k=1}^{30} u_{j k} \chi_{k}(t)$, where $\mathbf{u}_{i}=\left(u_{i 1}, \cdots, u_{i 30}\right)^{\prime}$ are the coefficient vectors.

In order to estimate a policy-specific $\beta$ curve, we can either introduce an interaction term ($Z\times S^*$) in the RHS of \eqref{fcr} or we can fit two separate models for $Y$ series separately, i.e
\begin{eqnarray}
Y^{(1)}_{j}(t)&=&f^{(1)}_{0}(t)+S^{*(1)}_j(t-1)\beta^{(1)}(t) +b^{(1)}_{j}(t)+\epsilon^{(1)}_{j}(t)\label{fcr:Model1}\\
Y^{(2)}_{j}(t)&=&f^{(2)}_{0}(t)+S^{*(2)}_j(t-1)\beta^{(2)}(t) +b^{(2)}_{j}(t)+\epsilon^{(2)}_{j}(t)\label{fcr:Mode2}
\end{eqnarray}
where $Y^{(1)}_j(.)$ is the series associated with $Z=0$ and $Y^{(1)}_j(.)$ is the series associated with $Z=1$. Evidently, coefficient curves $\beta^{(1)}(t)$ and $\beta^{(2)}(t)$ will capture the differential impact of policy intervention. Since \eqref{fcr} can handle unequally sampled time series \cite{leroux2018dynamic}, we have used, separately, models of $Y$ series to extract the policy-specific coefficient curves. The fitting procedure was carried out using the \texttt{R} package \textit{fcr}.  All of the estimated $\beta$ values can be found in the supplemental materials located in the file \verb|Complete_result_beta.csv|.

Recall, the $\beta$ curves obtained using functional regression does not directly yield $\beta_A$ and $\beta_S$ that we need to compute $\mathcal{R}_0$ in \eqref{Eq: R0}. Hence, a post-hoc estimate of the curves associated with these two parameters are generated using the formula $\beta(t)=w\times \beta_A(t)+(1-w)\times \beta_S(t)$ where $w$ is dependent on the proportion of asymptomatic cases in the population with a further constraint that population mean of $\beta_A$ curve associated with the regime $Z=1$ will be dominated by the population mean of $\beta_A$ curve associated with the regime $Z=0$ \cite{catching2021examining}.

\begin{figure}[htbp]
    \centering
     \stackinset{l}{53mm}{t}{1pt}{\textbf{(a)}}{\stackinset{r}{59mm}{t}{1pt}{\textbf{(b)}}{\stackinset{l}{53mm}{b}{48mm}{\textbf{(c)}}{\stackinset{r}{59mm}{b}{48mm}{\textbf{(d)}}{\includegraphics[width = 0.9\textwidth]{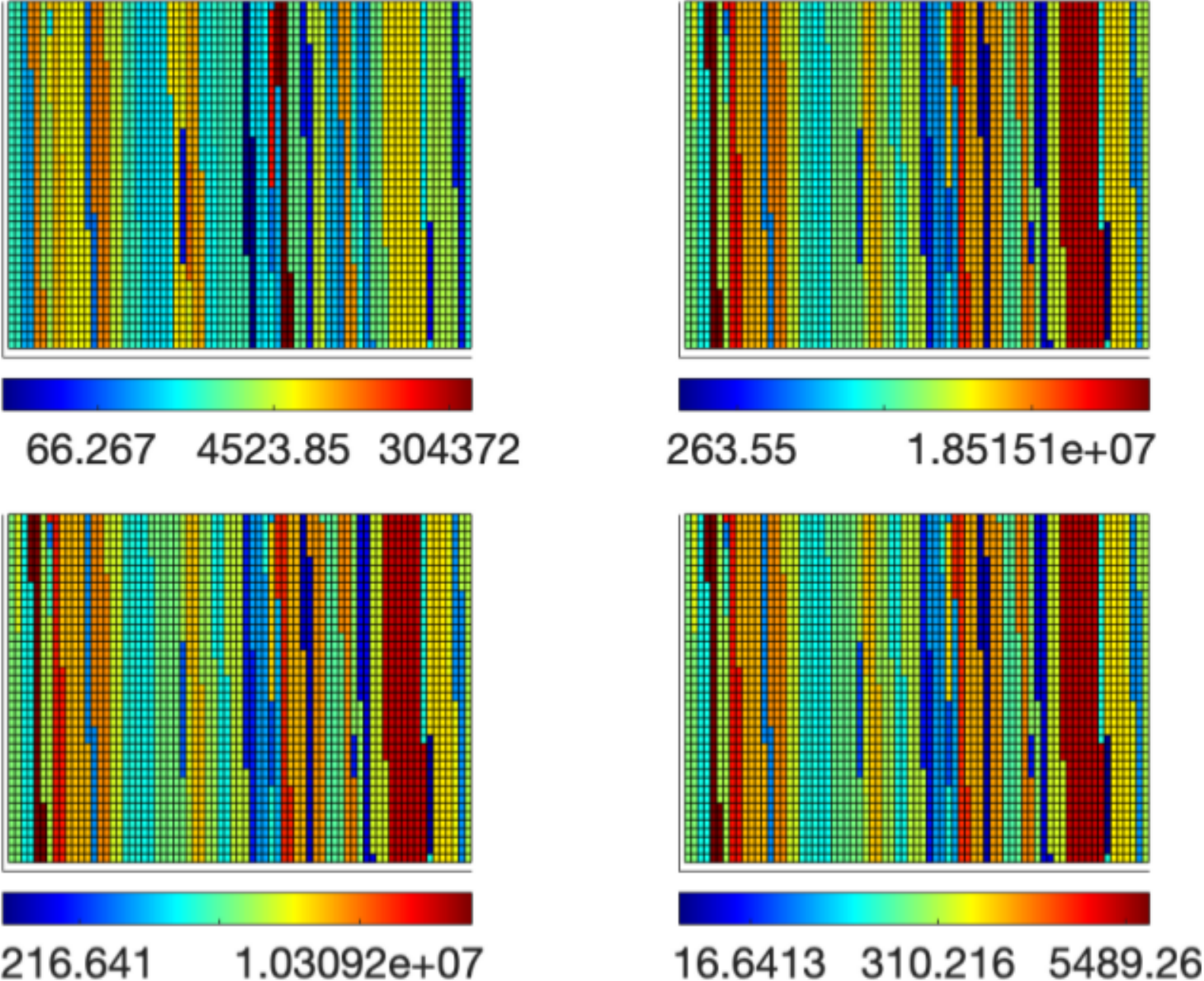}}}}}
    \caption{Simulation of the SIR model \eqref{Eq: Model0} with state-wise temporally evolving $\beta_A$ and $\beta_S$ with initial condition \textbf{(a)} taken from the $100$ day mark.  \textbf{(b)} - \textbf{(d)} simulated heat maps at the $110$, $130$, and $300$ day marks.  The colormap for the heat maps are on a $\log$-scale.}
    \label{Fig: Beta State}
\end{figure}

Now that we have obtained the time-evolving transmission parameters $\beta_A$ and $\beta_S$ for each state, we have another way to analyze the spatio-temporal dynamics of the disease.  In Sec. \ref{Sec: SIR} the redistribution function $\Psi$ was unity as the dynamics within the separate patches (counties) were the same.  With the varying $\beta$'s we have some non-trivial redistribution $u(x,y,t) = \Psi(x,y,t)I_s(t)$ that is implicitly dependent on $\beta_A$ and $\beta_S$.  First let us run our SIR model \eqref{Eq: Model0} with the new $beta$'s on a state-wide level.  The state-wide heat maps for the number of symptomatically infected individuals, $I_S$, are shown in Fig. \ref{Fig: Beta State} with the initial condition in \textbf{(a)} calculated at the $100$ day mark from the \url{usafacts.org} data set \cite{USAFacts}.  Fig. \ref{Fig: SIR}\textbf{(b - d)} show the solutions to the SIR model \eqref{Eq: Model0} with the varying $\beta$ values at day $110$, $130$, and $300$.  The colormap for the heat maps are on a $\log$-scale to highlight the large differences in disease proliferation between states, which is also what has been observed for the current pandemic.
In the heat maps the data for the states are taken in alphabetical order from left to right with each county represented down the column.  So the $\Omega$ state that uses data from California will take up multiple columns whereas the equivalent of Rhode Island will only take up three rows of a column.  

\begin{figure}[htbp]
    \centering
     \stackinset{l}{53mm}{t}{1pt}{\textbf{(a)}}{\stackinset{r}{60mm}{t}{1pt}{\textbf{(b)}}{\stackinset{l}{53mm}{b}{48mm}{\textbf{(c)}}{\stackinset{r}{60mm}{b}{48mm}{\textbf{(d)}}{\includegraphics[width = 0.9\textwidth]{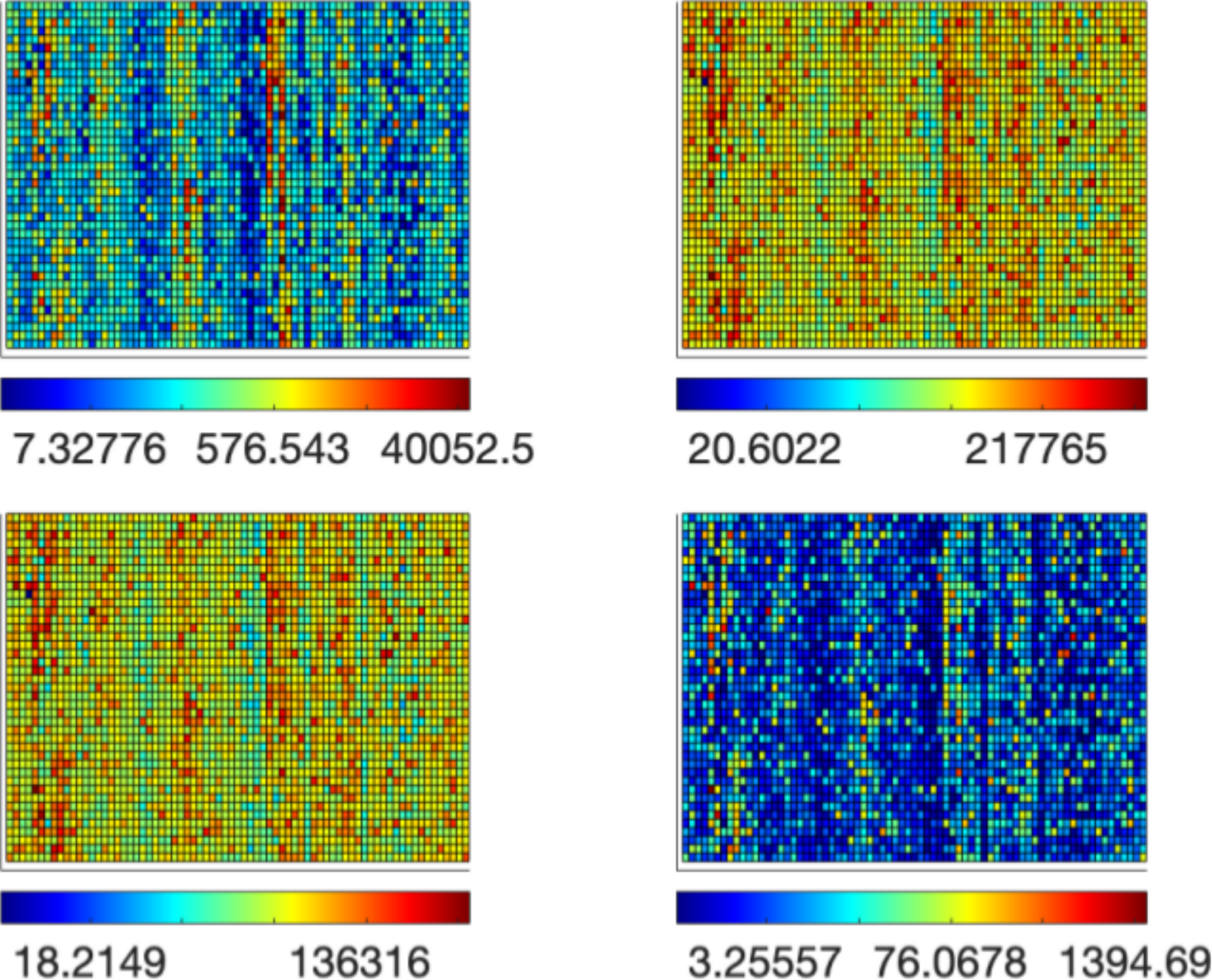}}}}}
    \caption{Simulation of the SIR model \eqref{Eq: Model0} with state-wise temporally evolving $\beta_A$ and $\beta_S$ with a county-wise downscaling through a simple product with the population distribution.  The initial condition \textbf{(a)} is taken from the $100$ day mark.  \textbf{(b)} - \textbf{(d)} simulated heat maps at the $110$, $130$, and $300$ day marks.  The colormap, with color bars on the bottom, for the heat maps are on a $\log$-scale.}
    \label{Fig: Beta County}
\end{figure}

While this is compelling, there is a lot of variance within individual states.  Although the $\beta$ values are state-wide we can downscale to the county level.  This downscaling can be done in many different ways by modifying the redistribution function $\Psi$.  For simplicity let us redistribute the state-wide results by taking the product with the population distribution.  This yields the county-wise number of symptomatically infected individuals, $I_S$, which is shown in Fig. \ref{Fig: Beta County}.  As before, the initial condition in \textbf{(a)} is calculated at the $100$ day mark from the \url{usafacts.org} data set \cite{USAFacts}.  Fig. \ref{Fig: SIR}\textbf{(b - d)} show the solutions to the SIR model \eqref{Eq: Model0} with the varying $\beta$ values at day $110$, $130$, and $300$.  The colormap, with color bars on the bottom, for the heat maps are on a $\log$-scale to highlight the large differences in disease proliferation between states, which is also what has been observed for the current pandemic.
Even with this very simple down-scaling we see a lot more structure in the heat maps.  This intensity structure is precisely what could help policy makers dictate location-specific policy and send resources where they are predicted to be needed the most.

\begin{figure}[htbp]
    \centering
    \includegraphics[width = 0.9\textwidth]{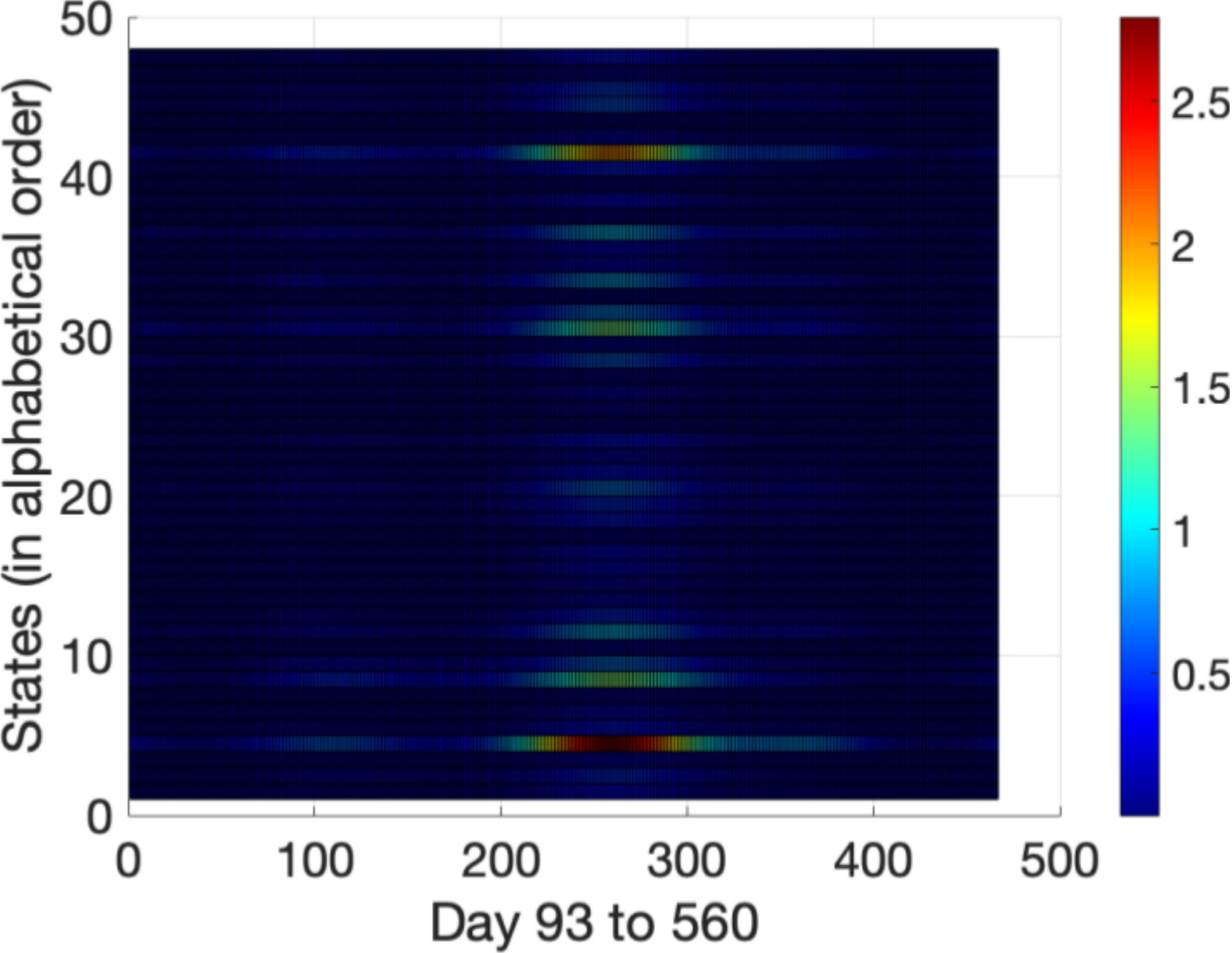}
    \caption{Heat map of the basic reproduction number, $\mathcal{R}_0$.  The 48 states of $\Omega$ are organized in alphabetical order from $1$ to $48$ (increasing ordinate direction).  The abscissa is marked from day $93$ to day $560$ corresponding to tick marks $0$ to $467$.  Finally, the colormap, with the color bar on the right, is on a linear scale from $0$ to $3$, which while calculated for a proof-of-concept toy model, is on par with that of the $\mathcal{R}_0$ of Covid-19.}
    \label{Fig: R0}
\end{figure}

As mentioned in Sec. \ref{Sec: SIR}, epidemiologists often estimate a basic reproduction number, $\mathcal{R}_0$, to indicate the overall course of a disease, and it is often used to inform policy decisions.  Now, that we have the $\beta$ values, we have the tools to calculate the spatio-temporally evolving $\mathcal{R}_0$.  We present a heat map of the $\mathcal{R}_0$ in Fig. \ref{Fig: R0}.  The 48 states of $\Omega$ are organized in alphabetical order from $1$ to $48$ (increasing ordinate direction).  The abscissa is marked from day $93$ to day $560$ corresponding to tick marks $0$ to $467$.  Finally, the colormap, with the color bar on the right, is on a linear scale from $0$ to $3$.  Notice that the basic reproduction number is similar to that of the current pandemic even though the calculation was done for a hypothetical country.  We also observe that as the pandemic carries on, the $\mathcal{R}_0$ increases, but then there is a sudden decrease.  This would be due to policy changes in the various states of $\Omega$, which to compare with our real-world situation, is most likely due to increased masking and locking down.

\section{Diffusion-driven redistribution between counties}
\label{Sec: PDE}

In Sec. \ref{Sec: SIR} we simulated spatio-temporal behavior in a purely dynamical manner by considering the individual counties as uniform and non-interacting, then Sec. \ref{Sec: Likelihood} considered non-uniformity in the form of temporally varying state-wide transmission parameters, $\beta_A$ and $\beta_S$.  One major issue of the current pandemic was the speed at which it spread across the world, effectively catching us by surprise.  We hypothesize that the speed of the spread was due to the significantly high rate of asymptomatic transmission compared to other similar diseases, which indicates the importance of considering interaction between locations.  Indeed we observed Covid-19 quickly spreading from one ``hot zone'' to another, however we only small indications of such hot zones in the simulations of the previous two sections.  

Suppose that highly asymptomatic diseases do not significantly reduce travel between locations since individuals may not even know they are carriers.  This is conducive to close interactions with other individuals thereby transmitting the disease even without the presence of symptoms.  Although there are many transport-dynamics models such as that of Brockmann \textit{et al.} \cite{Brockmann2006}, we notice that long distance travel reduced significantly and even halted once Covid-19 accelerated.  However, the hot zones kept evolving, which indicates many short distance interactions spatially propagating through the population, similar to Brownian-like motion.  Therefore, let us model the redistribution of asymptomatic individuals as a diffusive process, which will then implicitly (through the SIR model \eqref{Eq: Model0}) redistribute symptomatic individuals.

If we consider $I_A(t)$ in \eqref{Eq: Model0} as the ``stationary'' proliferation of asymptomatic infections and $\Psi(x,y, t = 0)$ as the standard population-wise distribution, then $u(x,y,t) = \Psi(x,y,t)I_A(t)$ is the redistributed proliferation.  Now we suppose the redistribution is diffusion-driven; i.e., $u(x,y,t)$ satisfies the diffusion equation
\begin{equation}
    \frac{\partial u}{\partial t} = \del \cdot \left(D(x,y)\del u\right);\qquad u(x,y,0) = \Psi(x,y,0)I_A(0),
    \label{Eq: PDE}
\end{equation}
where $D(x,y)$ is the diffusivity, which is proportional to the population, and the proliferation is driven by the SIR model \eqref{Eq: Model0}.  We set the diffusivity to be proportional to the population since the disease spreads faster in populations of higher densities; i.e., we expect the disease from a particular hot zone to spread to nearby places, but also eventually to far away places with a preference towards locations of high density.  Further, since we observed that Covid-19 started to spread from a particular hot zone once enough individuals were infected, and therefore we use a piecewise diffusivity
\begin{equation*}
    D(x,y) = \begin{cases}
    P(x,y)/\max(P) & \text{  if $u(x,y) > u_*$},\\
    0 & \text{  otherwise};
    \end{cases}
\end{equation*}
where $P(x,y)$ is the count-wise population and $u_*$ is the threshold for diffusivity.

While we model the redistribution as one physical process, it need not be a physical process at all, but rather a mathematical construct which implicitly houses the physical information of the system, in a possibly convoluted manner, such as with Schrodinger's equation.

Since we need to solve a partial differential equation (PDE) coupled to a system of ordinary differential equations (ODEs) with spatio-temporally varying parameters calculated through a statistical model, while also potentially hamstrung by geographic and data analytic considerations, the numerical methods could become quite complex.  In the interest of simplicity, we solve the SIR model \eqref{Eq: Model0} with \verb|ode45|, as done in Sec. \ref{Sec: SIR}, over a single day, and with single day temporal discretization we solve the PDE \eqref{Eq: PDE} using the Forward in Time Centered in Space method (FTCS).  We acknowledge that there are convergence and accuracy considerations, and that the numerics could be made far more sophisticated, and we leave that to a future study.  To guarantee convergence we take care to satisfy the Courant–Friedrichs–Lewy (CFL) condition \cite{CFL} by scaling the county-wise population by the population of the largest county to produce the diffusivity $D(x,y)$.  Even with the heavily simplified numerics, some of the simulations took several hours to run on an Early 2015 Macbook Pro.

\begin{figure}
    \centering
    \stackinset{l}{}{t}{3mm}{\textbf{(a)}}{\includegraphics[width = 0.45\textwidth]{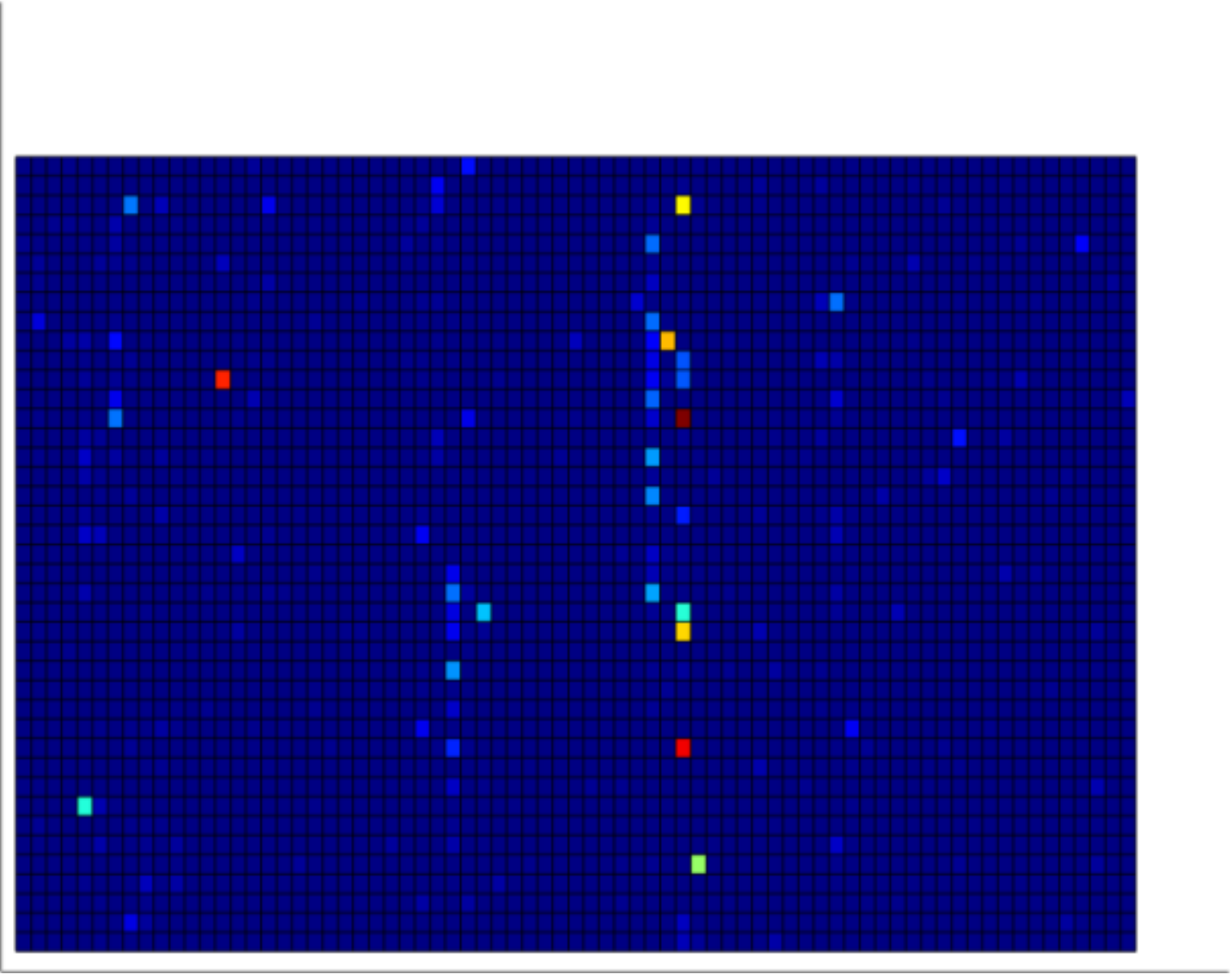}}\qquad
    \stackinset{l}{}{t}{3mm}{\textbf{(b)}}{\includegraphics[width = 0.45\textwidth]{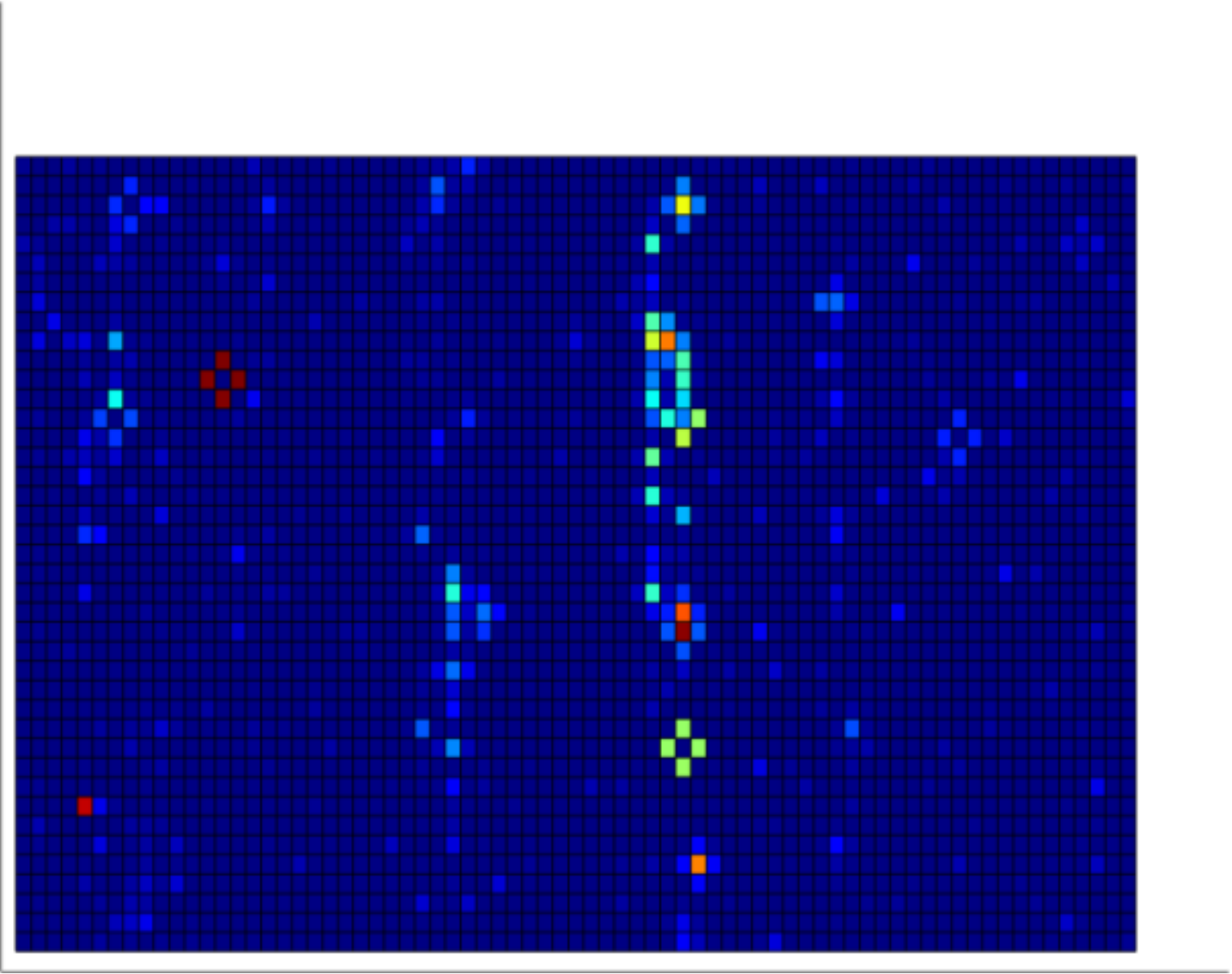}}
    \stackinset{l}{}{t}{3mm}{\textbf{(c)}}{\includegraphics[width = 0.45\textwidth]{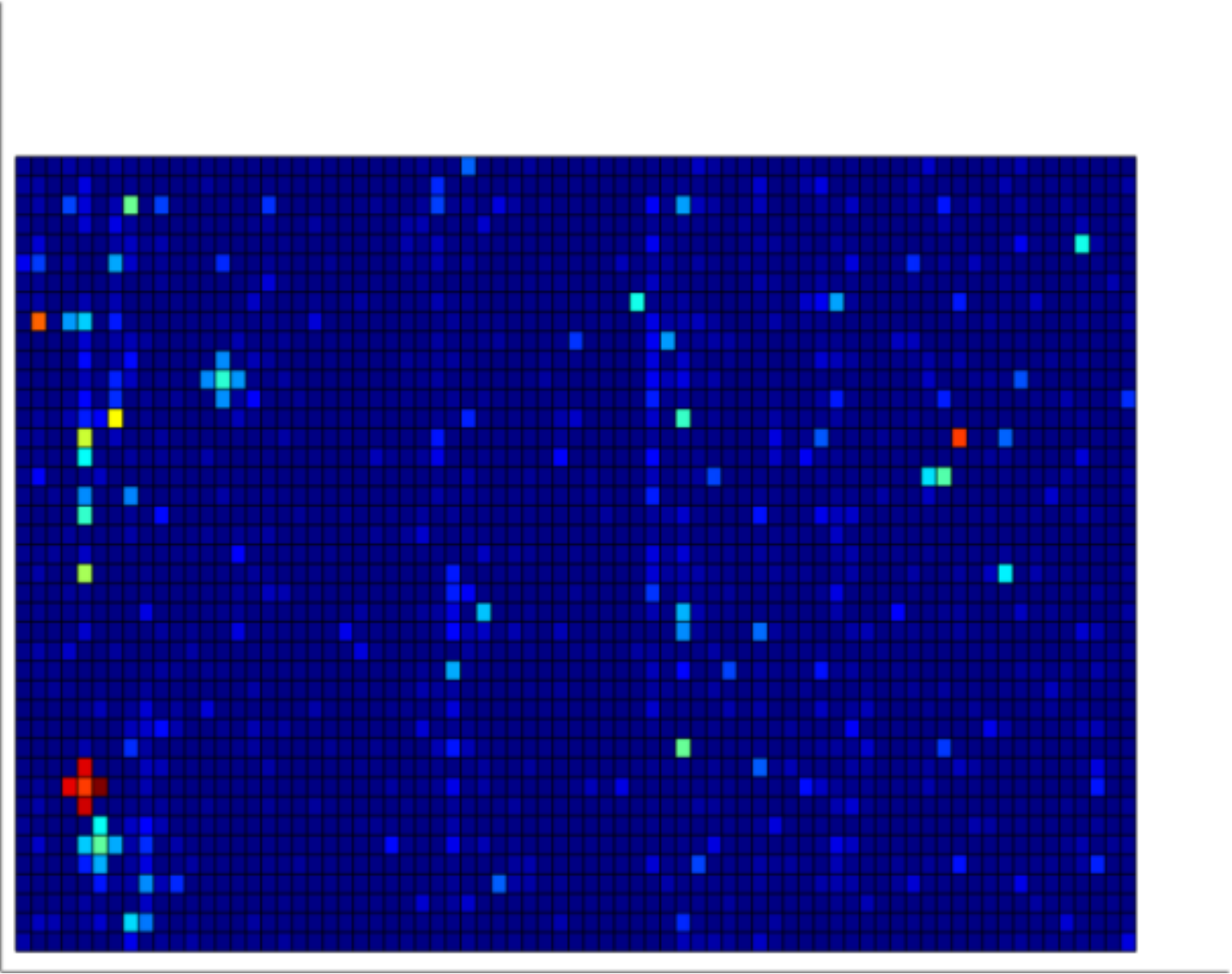}}\qquad
    \stackinset{l}{}{t}{3mm}{\textbf{(d)}}{\includegraphics[width = 0.45\textwidth]{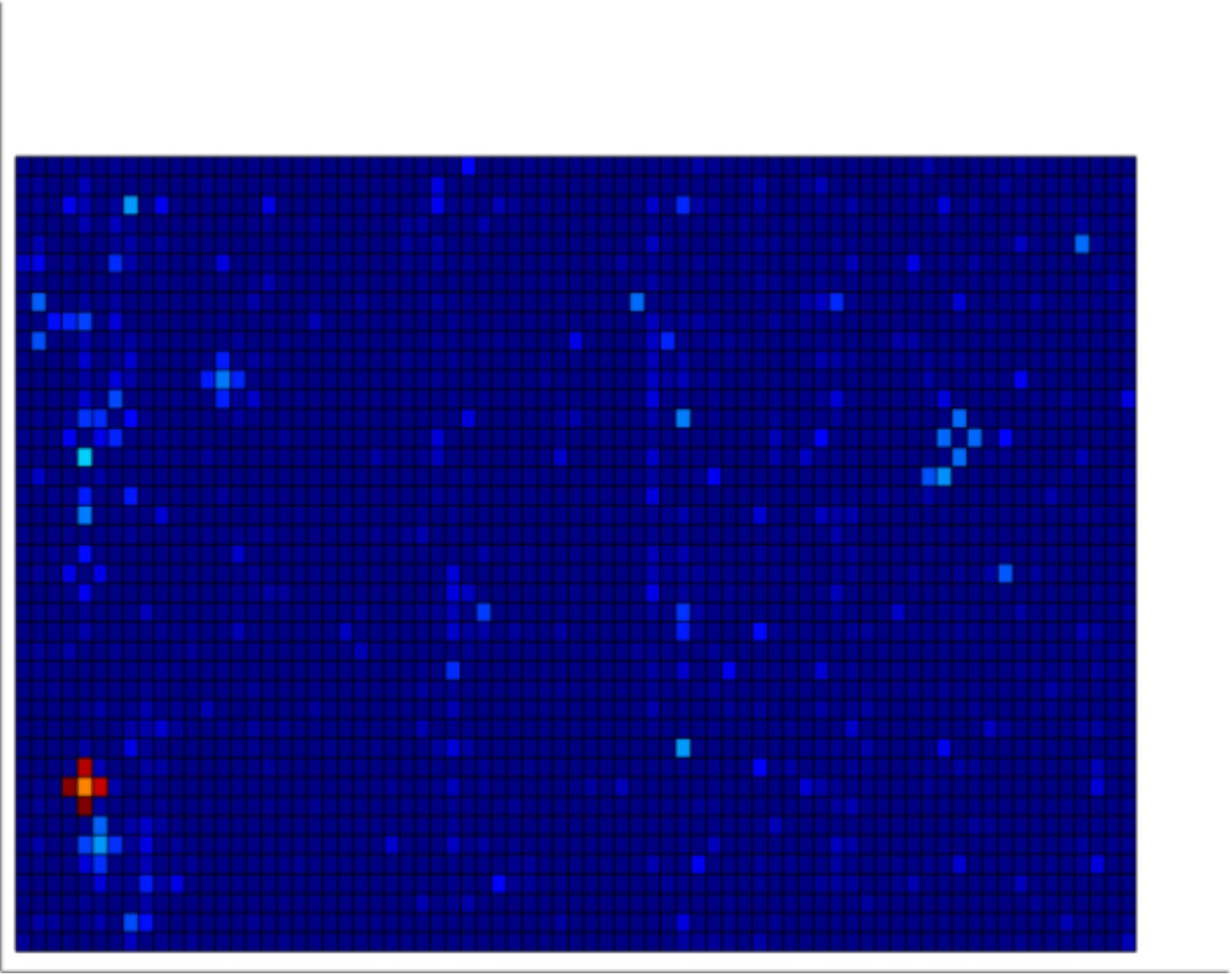}}
    \stackinset{l}{}{t}{3mm}{\textbf{(e)}}{\includegraphics[width = 0.45\textwidth]{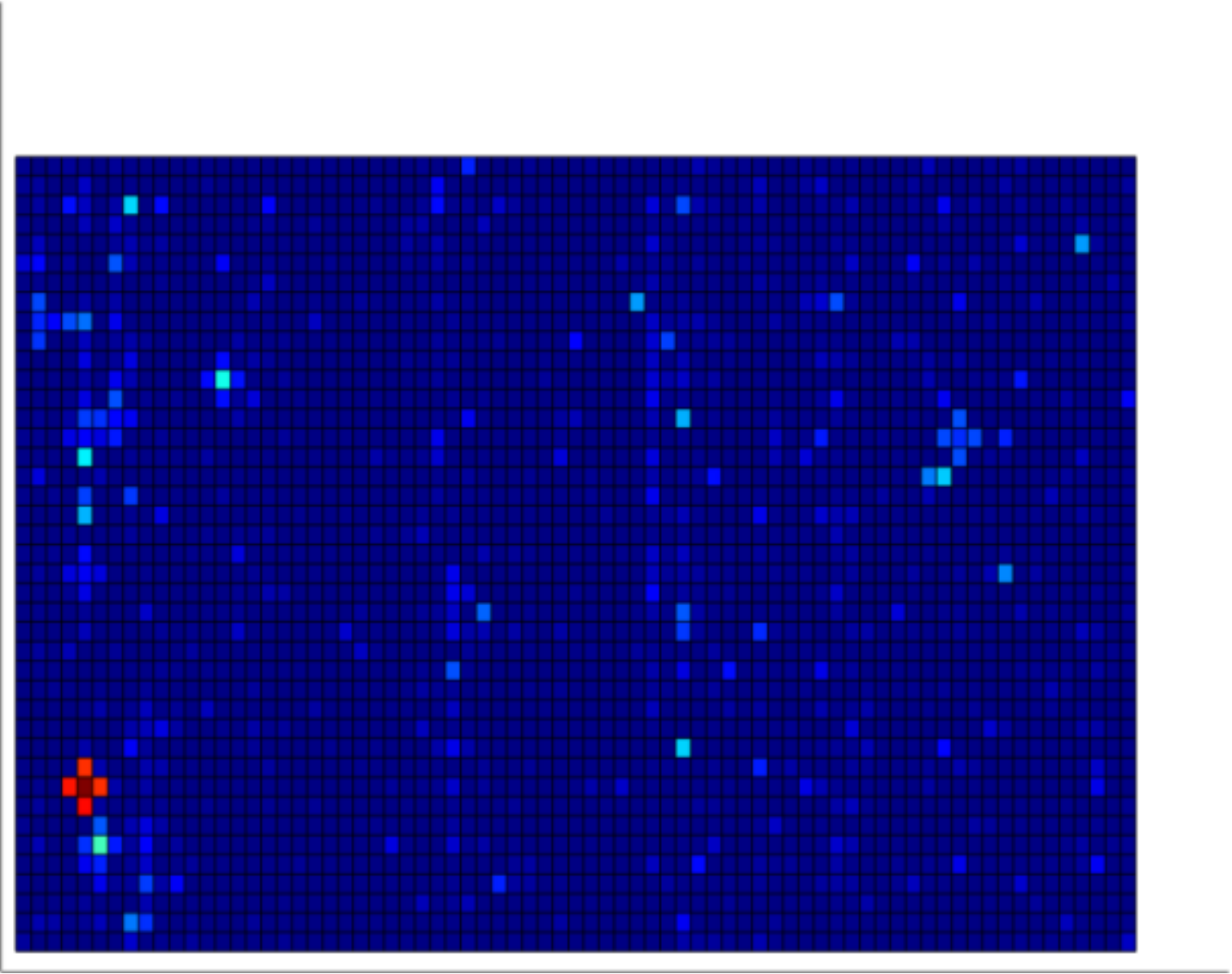}}\qquad
    \stackinset{l}{}{t}{3mm}{\textbf{(f)}}{\includegraphics[width = 0.45\textwidth]{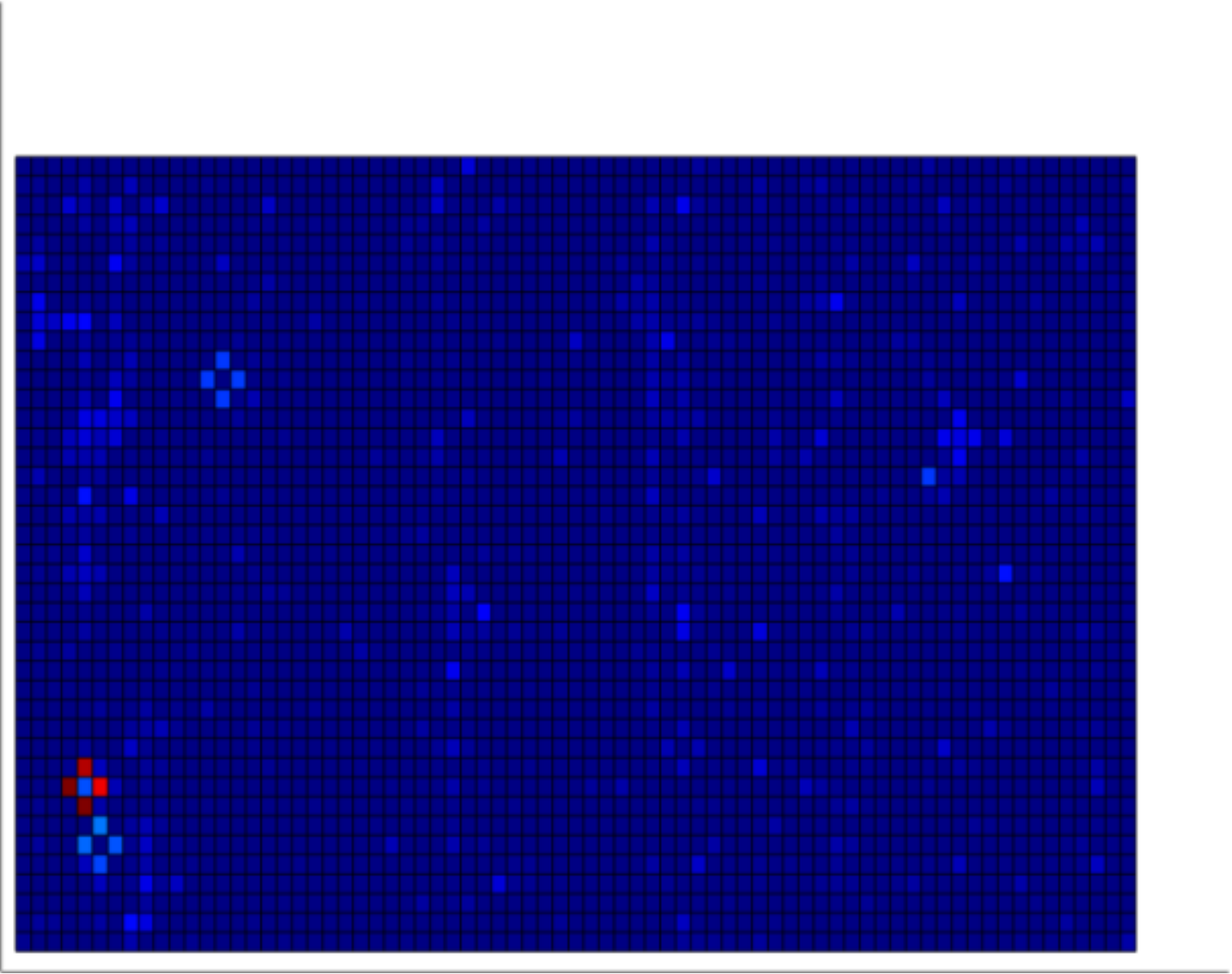}}
    \caption{Illustration of the numerical scheme to solve the PDE redistribution model \eqref{Eq: PDE} coupled with the SIR model \eqref{Eq: Model0} using FTCS and Runge-Kutta.  \textbf{(a) - (f)}  Subsequent progression of the scheme through three timesteps:  \textbf{(a, b)} $100$ day mark, \textbf{(c, d)} $101$ day mark, \textbf{(e, f)} $102$ day mark.  The heat maps in \textbf{(a, c, e)} are of the stationary proliferation with the standard distribution based on population, and \textbf{(b, d, f)} are post-diffusion distributions.  Since there was no way to accurately track asymptomatic cases, all of the heat maps are simulated.  The colormap is linear for the purposes of the demonstration.}
    \label{Fig: Numerics}
\end{figure}

We illustrate the numerical scheme in Fig. \ref{Fig: Numerics}.  We plot subsequent progression of the scheme from \textbf{(a, b)} the $100$ day mark to \textbf{(e, f)} the $102$ day mark.  The left hand column \textbf{(a, c, e)} represent the stationary proliferation with the standard distribution based on population, and the right hand column \textbf{(b, d, f)} are the post-diffusion distributions.  From \textbf{(a)} to \textbf{(b)} we observe the majority of the diffusion coming from counties with a large number of cases.  Once it diffuses, the disease continues to proliferate (\textbf{(b)} to \textbf{(c)}).  Then it diffuses more, then proliferates, and the process continues.  Of course, this is due to the discretization, and analytically the proliferation and redistribution would happen at the same time in a coupled manner.

\begin{figure}[htbp]
    \centering
     \stackinset{l}{56mm}{t}{1pt}{\textbf{(a)}}{\stackinset{r}{56mm}{t}{1pt}{\textbf{(b)}}{\stackinset{l}{56mm}{b}{50mm}{\textbf{(c)}}{\stackinset{r}{56mm}{b}{50mm}{\textbf{(d)}}{\includegraphics[width = 0.9\textwidth]{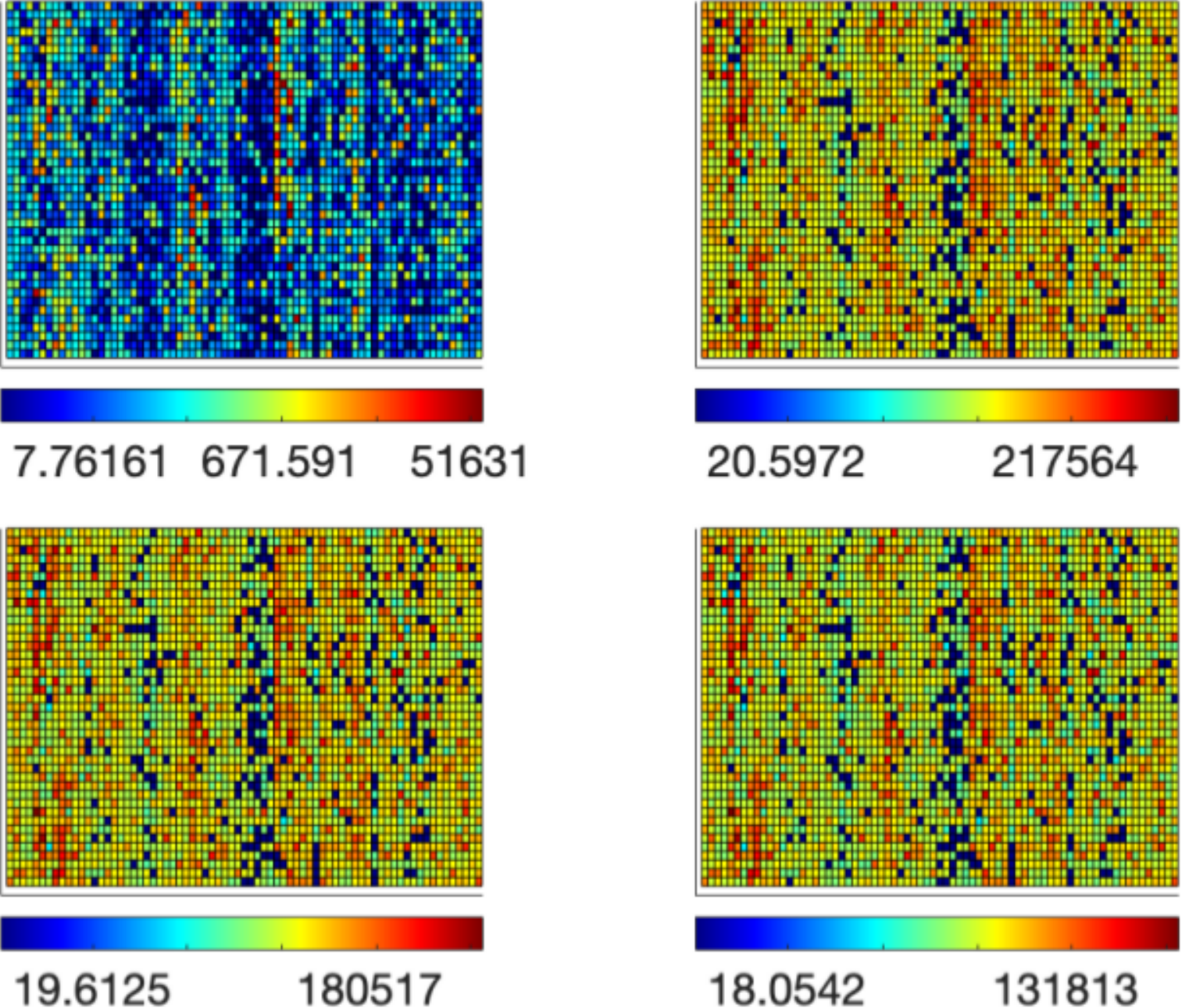}}}}}
    \caption{Simulation of the symptomatic cases by solving the PDE redistribution model \eqref{Eq: PDE} coupled with the SIR model \eqref{Eq: Model0} with state-wise temporally evolving $\beta_A$ and $\beta_S$.  The initial condition \textbf{(a)} is taken from the $100$ day mark.  \textbf{(b)} - \textbf{(d)} simulated heat maps at the $110$, $120$, and $130$ day marks.  The colormap, with color bars on the bottom, for the heat maps are on a $\log$-scale.}
    \label{Fig: PDE Symp}
\end{figure}

Similar to those of the previous sections, we present the simulation the symptomatic cases by solving \eqref{Eq: PDE} coupled with \eqref{Eq: Model0} using the $\beta$ values of Sec. \ref{Sec: Likelihood} in Fig. \ref{Fig: PDE Symp}.  The initial condition \textbf{(a)} is taken from the $100$ day mark, and then \textbf{(b)} - \textbf{(d)} are simulated at the $110$, $120$, $130$ day marks.  The colormap, with color bars on the bottom, for the heat maps are on a $\log$-scale to highlight the large differences in disease proliferation between counties.  It was necessary to shorten the prediction window because of computational expenditure, which a more sophisticated numerical scheme, as acknowledged in the previous paragraph, would rectify.  Despite these limitations, we see the hot zones more clearly in the heat maps.

\begin{figure}[htbp]
    \centering
     \stackinset{l}{56mm}{t}{1pt}{\textbf{(a)}}{\stackinset{r}{57mm}{t}{1pt}{\textbf{(b)}}{\stackinset{l}{56mm}{b}{49mm}{\textbf{(c)}}{\stackinset{r}{57mm}{b}{49mm}{\textbf{(d)}}{\includegraphics[width = 0.9\textwidth]{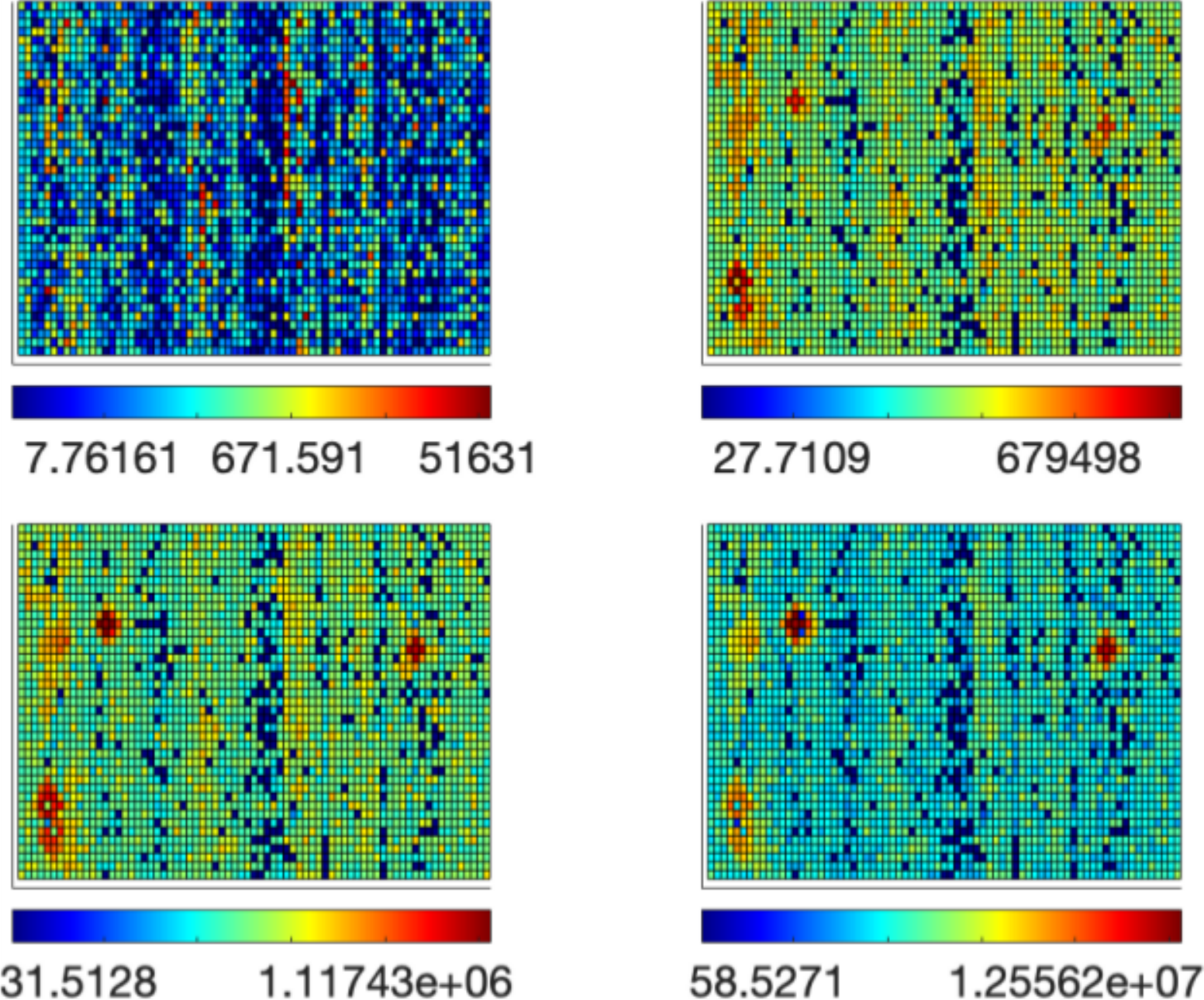}}}}}
    \caption{Simulation of asymptomatic cases by solving the PDE redistribution model \eqref{Eq: PDE} coupled with the SIR model \eqref{Eq: Model0} with state-wise temporally evolving $\beta_A$ and $\beta_S$.  \textbf{(a)} - \textbf{(d)} simulated heat maps at the $100$, $110$, $120$, and $130$ day marks.  The colormap, with color bars on the bottom, for the heat maps are on a $\log$-scale.}
    \label{Fig: PDE Asymp}
\end{figure}

Although the asymptomatic cases were presented in Fig. \ref{Fig: Numerics}, it is perhaps more compelling to get the full picture of the evolution of asymptomatic cases, and hence present the evolution in Fig. \ref{Fig: PDE Asymp}.  Just like in Fig. \ref{Fig: PDE Symp} we solve the PDE redistribution model \eqref{Eq: PDE} coupled with the SIR model \eqref{Eq: Model0} with state-wise temporally evolving $\beta_A$ and $\beta_S$.  Once again, the initial condition is taken from the $100$ day mark from the \url{usafacts.org} data set \cite{USAFacts}.  The $110$, $120$, and $130$ day marks are simulated in \textbf{(b)} - \textbf{(d)}.  The colormap, with color bars on the bottom, for the heat maps are on a $\log$-scale to highlight the large differences in disease proliferation between counties, which will make the hot zones clearer.
Here we see a quite compelling illustration of the hot zones, and even their evolution.  Hypothetically, if there is a disease where the symptomatic cases lag behind the asymptomatic cases, and we do not have an accurate way of tracking asympotmatic cases, a model such as the one presented here could indicate where resources may be necessary before a hot zone arises and the healthcare system becomes overloaded.

It should be noted that in Fig. \ref{Fig: PDE Asymp} the case numbers may seem unrealistic, but the reader is reminded that this is simply a proof-of-concept toy model to demonstrate that such a modelling framework can predict and hopefully help mitigate a disaster when applied to a real-world situation.

\section{Discussion and Future Work}
\label{Sec: Conclusion}


Although infectious disease has been studied for centuries, and at an especially accelerated rate in modern times, the current Covid-19 pandemic has revealed that there is still a lot we do not know about infectious disease.  In particular, the highly asymptomatic nature of the disease caught the world by surprise.  Our globalized society creates an environment conducive to the swift spread of diseases over long distances.  The ability for the disease to cause an unbearable load on local medical infrastructure, and then move unpredictably to another hot zone, created problems of resource allocation.  Even when the resources were available to release the pressure on our infrastructure, they did not always get to where it was needed.  It is evident that more spatio-temporal models for infectious disease will be necessary in the future.

In this investigation, we progress through three types of spatio-temporal models for highly asymptomatic diseases.  We first derive an SIR-like model \eqref{Eq: Model0} with fixed parameter values in Sec. \ref{Sec: SIR}.  The model is applied at the county-wise level where all the counties are assumed to be uniform and non-interacting, which reveals some spatial structure, but misses more subtle structure, such as hot zones.  Then in Sec. \ref{Sec: Likelihood} we derive the state-wide spatio-temporal evolution of the transmission parameters, $\beta_A$ and $\beta_S$, through likelihood based estimation.  The variance in the transmission shows the effect of behavioral and policy changes, such as masking and lockdowns.  This is particularly revealing in the evolution of the basic reproduction number, $\mathcal{R}_0$, \eqref{Eq: R0} in Fig. \ref{Fig: R0}.  Finally, in Sec. \ref{Sec: PDE} we derive a complete (including interactions between regions) spatio-temporal model by coupling a PDE governing redistribution \eqref{Eq: PDE} to the SIR model \eqref{Eq: Model0} with the state-wide spatio-temporally varying transmission parameters.  This is where we see hot zones appearing and evolving, which strengthens the modeling framework as a proof-of-concept.

We foresee many more future projects arising from the modeling framework presented here.  For example, notice that even at the state level population density is the strongest indicator of disease proliferation, regardless of clearly beneficial policy implementation such as masking.  So does masking not work?  There is an abundance of articles that show masking is effective (e.g., \cite{MaskingNature2020, ChengMasking2021, BrooksButlerMasking2021}).  One way to see this is by comparing predicted results for no mitigation policies versus the location by location differences in the actual case counts due to the implemented mitigation techniques in those areas.  While, this was not included in the manuscript, it shows the importance of spatio-temporal models for disease transmission, and as more data is produced and more sophisticated models are derived, the possibility for such comparisons becomes very promising.

Moreover, the numerical methods in Sec. \ref{Sec: PDE} can be improved significantly.  We propose a future study with rigorous numerical analysis on solving the coupled PDE-SIR model.  Most importantly, the SIR ODE system and the PDE should be discretized together.  Runge-Kutta can still be used to integrate the ODE and Crank-Nicolson \cite{CrankNicolson} can be used to integrate the PDE, which should be done in unison over one timestep.  Crank-Nicolson was avoided (in favor of FTCS) due to the 2-dimensional nature of the problem.  While the matrix equation can certainly be written down, the pentadiagonal system would be impossible to solve in any practical sense even for a highly optimized algorithm such as that of Strassen \cite{Strassen1969}.  We would need to employ an alternating-direction implicit method, similar to that of Peaceman and Rachford Jr. \cite{PeacemanRachford1955}, in order to avoid being hamstrung by the computational expense.  Even then, the computational complexity would be quite high, and we foresee the need to parallelize several of the processes to run on a supercomputing cluster.

A major test for the modeling framework will be its performance when applied to real-world geographical constraints.  At any other time the data for such a problem would not be available.  The unique circumstances of the current pandemic provides us with multiple years of epidemiological data.  By carefully constructing numerical methods for the highly non-uniform grid formed by the counties of the United States, we would have direct comparisons between the simulations and the data.  Section \ref{Sec: PDE} also only considers the diffusion of asymptomatic cases, but there are other factors that could affect the deviation of the distribution of cases from that of the standard population-wise distribution.  In future studies, we need not be beholden to a physical process, but rather use the redistribution $\Psi$ as a mathematical construct similar to the way Schrodinger's equation is used in Quantum Mechanics.  

Finally, it is not unlikely that the ebb and flow of such a complex system is chaotic, especially as the pandemic has carried on for so long and the virus has mutated several times.  As with the weather, due to errors in measurement (no matter how small) we can only hope to reproduce the qualitative long term behavior rather than precise long term predictions.  However, we may be able to reliably forecast a week or a month ahead, which is effective for weather-related preparations, and it may be effective for infectious disease as well.  Further, the manuscript does not consider mutations or vaccination, but as we are currently observing, this pandemic is uniquely complex due to the rate at which SARS CoV2 mutates, which occasionally evades vaccination.  We foresee the inclusion of mutations and vaccinations in future studies as an important step in understanding the spatio-temporal evolution of a highly contagious disease.

\section*{Acknowledgments}
A.R. appreciate the support of the Amath department at UW,  A.P. appreciates the support of the Math and Stats department at TTU, and R.K. and S.G. appreciate the support of the Stats department at UNL.

\bibliographystyle{unsrt}
\bibliography{Refs}

\begin{thebibliography}{10}

\bibitem{Gurarie2013EncounterRates}
Eliezer Gurarie and Otso Ovaskainen.
\newblock Towards a general formalization of encounter rates in ecology.
\newblock {\em Theor. Ecol.}, 6:189--202, 2013.

\bibitem{lloyd2004spatiotemporal}
Alun~L Lloyd and Vincent~AA Jansen.
\newblock Spatiotemporal dynamics of epidemics: synchrony in metapopulation
  models.
\newblock {\em Mathematical biosciences}, 188(1-2):1--16, 2004.

\bibitem{Brockmann2006}
D.~Brockmann, L.~Hufnagel, and T.~Geisel.
\newblock The scaling laws of human travel.
\newblock {\em Nature}, 439:462--465, 2006.

\bibitem{StatisticalMechanics}
A.~I. Khinchin.
\newblock {\em Statistical Mechanics}.
\newblock Dover Publications, Inc., 1949.

\bibitem{Adam2020}
David Adam.
\newblock Modelers struggle to predict the future of the covid-19 pandemic.
\newblock {\em The Scientist}, 2020.

\bibitem{BertozziChallenges20}
A.~Bertozzi, E.~Franco, G.~Mohler, M.~B. Short, and D.~Sledge.
\newblock The challenges of modeling and forecasting the spread of covid-19.
\newblock {\em Proc. Nat. Acad. Sci.}, 117(29):16732--16738, 2020.

\bibitem{CovidViralShedding}
J.~Y. Noh, J.~G. Yoon, H.~Seong, W.~S. Choi, J.~W. Sohn, W.~J. Kim, and J.~Y.
  Song.
\newblock Asymptomatic infection and atypical manifestations of covid-19:
  Comparison of viral shedding duration.
\newblock {\em J. Infect.}, 81(5):816--846, 2020.

\bibitem{CovidMetaAnalysis}
M.~Alene, L.~Yismaw, M.~A. Assemie, D.~B. Katema, B.~Mengist, B.~Kassie, and
  T.~Y. Birhan.
\newblock Magnitude of asymptomatic covid-19 cases throughout the course of
  infection: A systematic review and meta-analysis.
\newblock {\em PLOS ONE}, page 0249090, 2021.

\bibitem{UsherwoodCovid2021}
T.~Usherwood, Z.~LaJoie, and V.~Srivastava.
\newblock A model and predictions for covid-19 considering population behavior
  and vaccination.
\newblock {\em Sci. Rep.}, 11:12051, 2021.

\bibitem{McAloonCovid2020}
C.~McAloon, A.~Collins, K.~Hunt, A.~Barber, A.~W. Byrne, F.~Butler, M.~Casey,
  J.~Griffin, E.~Lane, D.~McEvoy, P.~Wall, Green M., L.~O'Grady, and S.~J.
  More.
\newblock Incubation period of covid-19: a rapid systematic review and
  meta-analysis of observational research.
\newblock {\em Epidemiology}, 10(8):e039652, 2020.

\bibitem{USAFacts}
Us covid-19 cases and deaths by state, 2020.

\bibitem{Runge}
C.~D.~T. Runge.
\newblock \"{U}ber die numerische aufl\"{o}sung von differentialgleichungen.
\newblock {\em Mathematische Annalen}, 46(2):167--178, 1895.

\bibitem{Kutta}
M.~Kutta.
\newblock Beitrag zur n\"{a}herungsweisen integration totaler
  differentialgleichungen.
\newblock {\em Zeitschrift f\"{u}r Mathematik und Physik}, 46:435--453, 1901.

\bibitem{diekmann2010construction}
Odo Diekmann, JAP Heesterbeek, and Michael~G Roberts.
\newblock The construction of next-generation matrices for compartmental
  epidemic models.
\newblock {\em Journal of the Royal Society Interface}, 7(47):873--885, 2010.

\bibitem{van2008spatial}
P~Van~den Driessche.
\newblock Spatial structure: Patch models.
\newblock In {\em Mathematical epidemiology}, pages 179--189. Springer, 2008.

\bibitem{illian2008statistical}
Janine Illian, Antti Penttinen, Helga Stoyan, and Dietrich Stoyan.
\newblock {\em Statistical analysis and modelling of spatial point patterns},
  volume~70.
\newblock John Wiley \& Sons, 2008.

\bibitem{bruno2009simple}
Francesca Bruno, Peter Guttorp, Paul~D Sampson, and Daniela Cocchi.
\newblock A simple non-separable, non-stationary spatiotemporal model for
  ozone.
\newblock {\em Environmental and ecological statistics}, 16(4):515--529, 2009.

\bibitem{rodrigues2010class}
Alexandre Rodrigues and Peter~J Diggle.
\newblock A class of convolution-based models for spatio-temporal processes
  with non-separable covariance structure.
\newblock {\em Scandinavian Journal of Statistics}, 37(4):553--567, 2010.

\bibitem{leroux2018dynamic}
A.~Leroux, L.~Xiao, C.~Crainiceanu, and W.~Checkley.
\newblock Dynamic prediction in functional concurrent regression with an
  application to child growth.
\newblock {\em Statistics in Medicine}, 37(8):1376--1388, 2017.

\bibitem{catching2021examining}
A.~Catching, S.~Capponi, M.~T. Yeh, S.~Bianco, and R.~Andino.
\newblock Examining the interplay between face mask usage, asymptomatic
  transmission, and social distancing on the spread of covid-19.
\newblock {\em Sci. Rep.}, 11:15998, 2021.

\bibitem{CFL}
R.~Courant, K.~Friedrichs, and H.~Lewy.
\newblock \"{U}ber die partiellen differenzengleichungen der mathematischen
  physik.
\newblock {\em Mathematische Annalen}, 100:32--74, 1928.

\bibitem{MaskingNature2020}
L.~Peeples.
\newblock Face masks: what the data say.
\newblock {\em Nature}, 586:186--189, 2020.

\bibitem{ChengMasking2021}
Y.~Cheng, N.~Ma, C.~Witt, S.~Rapp, P.~S. Wild, M.~O. Andreae, U.~P\"{o}schl,
  and H.~Su.
\newblock Face masks effectively limit the probability of sars-cov-2
  transmission.
\newblock {\em Science}, 372(6549):1439--1443, 2021.

\bibitem{BrooksButlerMasking2021}
J.~T. Brooks and J.~C. Butler.
\newblock Effectiveness of mask wearing to control community spread of
  sars-cov-2.
\newblock {\em JAMA Insights}, 325(10):998--999, 2021.

\bibitem{CrankNicolson}
J.~Crank and P.~Nicolson.
\newblock A practical method for numerical evaluation of solutions of partial
  differential equations of the heat conduction type.
\newblock {\em Proc. Camb. Phil. Soc.}, 43(1):50--67, 1947.

\bibitem{Strassen1969}
V.~Strassen.
\newblock Gaussian elimination is not optimal.
\newblock {\em Numer. Math}, 13(4):354--356, 1969.

\bibitem{PeacemanRachford1955}
D.~W. Peaceman and H.~H. Rachford~Jr.
\newblock The numerical solution of parabolic and elliptic differential
  equations.
\newblock {\em SIAM J. Appl. Math.}, 3(1):28--41, 1955.

\end{thebibliography}

\end{document}